\documentclass [12pt,twoside]{article}           
\usepackage[left,modulo]{lineno}
\usepackage{setspace}

\countdef\arxiv=201

\ifnum\arxiv=0 \input cpreb.tex \fi

\ifnum\arxiv=1

\usepackage{epsfig,times,lscape}
\usepackage[usenames]{color}
\usepackage{textcomp}  
    \ifnum\count102=1

    \fi

    \ifnum\count102=2
\fi

\newcommand{\nc}{\newcommand}

     \ifnum\count101=1
\nc{\qI}[1]{\section{{#1}}}
\nc{\qA}[1]{\subsection{{#1}}}
\nc{\qun}[1]{\subsubsection{{#1}}}
\nc{\qa}[1]{\paragraph{{#1}}}

\def\qpar{\vskip 2mm plus 0.2mm minus 0.2mm}
\def\qL{\hfill \break}
     \fi 

      \ifnum\count101=2
 \nc{\qI}[1]{\parindent=0mm \vskip 8mm 
{\centerline{\LARGE \color{red}#1}}\vskip 3mm}
\nc{\qA}[1]{\vskip 2.5mm \noindent 
{{\bf\large\color{blue}  #1}} \vskip 1mm \parindent=0mm}
 \nc{\qun}[1]{\vskip 1mm \noindent {\sl #1 }\quad }

\def\qL{\hfill \break}
\def\qpar{\vskip 2mm plus 0.2mm minus 0.2mm}

      \fi

\def\qth{\vrule height 12pt depth 0pt width 0pt}
\def\qtb{\vrule height 0pt depth 5pt width 0pt}

\nc{\qfoot}[1]{\footnote{{#1}}}

\newcommand{\promille}{
  \relax\ifmmode\promillezeichen
        \else\leavevmode\(\mathsurround=0pt\promillezeichen\)\fi}
\newcommand{\promillezeichen}{%
  \kern-.05em%
  \raise.5ex\hbox{\the\scriptfont0 0}%
  \kern-.15em/\kern-.15em%
  \lower.25ex\hbox{\the\scriptfont0 00}}

      \ifnum\count101=1
\def\qbu{\hfill \par \hskip 6mm $ \bullet $ \hskip 2mm}
\def\qee#1{\hfill \par \hskip 6mm (#1) \hskip 2 mm}
      \fi
      \ifnum\count101=2
\def\qbu{\hfill \par \hskip 4mm $ \bullet $ \hskip 2mm}
\def\qee#1{\hfill \par \hskip 4mm (#1) \hskip 2 mm}
      \fi

\def\qparr{ \vskip 1.0mm plus 0.2mm minus 0.2mm \hangindent=10mm
\hangafter=1}

     \ifnum\count101=1 
 
     \fi
     \ifnum\count101=2

  \def\qcitb#1{\noindent \hbox to 102mm{\hfill \small #1} \vskip 1mm}
      \fi

 \def\qpages#1{\count102=0{\loop\advance\count102 by 1
 \null \vfill\eject \ifnum\count102<#1 \repeat}}

\def\ql{\ \hbox{--}\ }

\def\qth{\vrule height 12pt depth 0pt width 0pt}
\def\qtb{\vrule height 0pt depth 5pt width 0pt}

\def\qv{\vskip 0.1mm plus 0.05mm minus 0.05mm}
\def\qhu{\hskip 0.6mm}
\def\qhv{\hskip 3mm}

\def\qhw{\hskip 1.5mm}
\def\qleg#1#2#3{\noindent {\bf \small #1\qhw}{\small #2\qhw}{\it \small #3}\qv }
\fi

\begin{document}
\thispagestyle{empty}



\markboth{{\sl \hfill  \hfill \protect\phantom{3}}}
        {{\protect\phantom{3}\sl \hfill  \hfill}}

\color{yellow} 
\hrule height 10mm depth 10mm width 170mm 
\color{black}

\vskip -12mm 

\centerline{\bf \large The physics of large-scale food crises}
\vskip 15mm

\centerline{\large 
Peter Richmond$ ^1 $, Bertrand M. Roehner$ ^2 $ and
Qing-hai Wang$ ^3 $}

\vskip 10mm
\large

%
{\bf \color{red} Abstract}\qL
Investigating the ``physics'' of food crises consists in
identifying features which are common to all large-scale
food crises. One element which stands out is the fact
that during a food crisis there is not only a surge in
deaths but also a correlative temporary decline in 
conceptions and subsequent births.
As a matter of fact, birth reduction may even start
several months before the death surge and can
therefore serve as an early warning signal of an impending
crisis. \qL
This scenario is studied in three cases of large-scale 
food crises.
Finland (1868), India (1867--1907), China (1960--1961).
It turns out that between the regional
amplitudes of death spikes and 
birth troughs there is a power law relationship.
This confirms what was already observed 
for the epidemic of 1918 in the United States
(Richmond et al. 2018b).\qL
In a second part of the paper we explain
how this relationship can be used
for the investigation of mass-mortality episodes in cases
where direct death data are either uncertain or nonexistent.

\vskip 2mm
\centerline{\it \small Version of 8 July  2018}
\vskip 2mm

{\small Key-words: Famine, death rate, birth rate, excess-death.
Bertillon effect}

\vskip 5mm

{\normalsize
1: School of Physics, Trinity College Dublin, Ireland.
Email: peter\_richmond@ymail.com \qL
2: Institute for Theoretical and High Energy Physics (LPTHE),
University Pierre and Marie Curie (Sorbonne), Paris, France and 
CNRS (National Center for Scientific Research).\qL
Email: roehner@lpthe.jussieu.fr.\qL
3: Physics Department, National University of Singapore.
Email: qhwang@nus.edu.sg
}

\vfill\eject

\qI{Introduction}

\qA{A plea for comparative analysis}

The
title of the present paper%
\qfoot{The paper was written by three physicists.
However,
we feel that our approach is very much in the
spirit of the methodological
guidelines defined by the French sociologist Emile
Durkheim (1898). Incidentally,
the same comparative approach was also
used in our
previously published papers in biodemography
(http://www.lpthe.jussieu.fr/~roehner/biodemo.html.)}
was inspired by a book
published seven years ago (Viswanathan et al. 2011)
under the title: ``The physics of foraging''.
From
foraging, i.e. the collection of food by animals,
to food crises there is of course a smooth transition. 
However, in this title it is the mention of the word
``physics'' which was
for us the most important source of inspiration for it
means that the
purpose of the book was to examine {\it basic features}
of foraging that are shared by many species.
To find common mechanisms in seemingly different
phenomena has been a permanent objective of physics
throughout its development over past centuries. 
Is there a common factor in
the fall of apples, rain drops, meteorites and the
``fall'' of the Moon toward the Earth?
We now know that
the common factor is the gravitational attraction.
Here, we will be guided by a similar agenda.
\qpar

It can be obsereved that such a comparative 
approach was quite
common in the late 19th century; see for instance
the works of Louis-Alphonse Bertillon (1872),
Alfred Espinas (1878), Jacques Bertillon (1892),
Emile Durkheim (1895). One may deplore that in recent decades
comparative studies of this kind
(among which we do not include meta-analyses
which is something different) have become rare.

\qA{The Bertillon effect:
from heat-wave mortality to large-scale food crises}

Here we will focus on the
Bertillon effect (Bertillon 1892, Richmond et al. 2018a,b)
which consists in the fact that any mass mortality
is followed 9 months later by a birth rate trough%
\qfoot{Note that the birth rate trough is
much larger (usually about 10 times larger) than the reduction
in births due to the fatalities. This was shown in detail in
Richmond et al. (2018a) 
and can also be seen from the simple fact
that excess fatalities would result in a 
one sided Heaviside fall not in a symmetrical and
fairly narrow trough. This trough is mostly due to a temporary 
reduction in the
conception rate among the survivors.}%
.
\qpar

Our previous parallel with gravitation becomes particularly relevant
here for indeed, just like gravity, the Bertillon effect has
a broad range of applicability. There is a long chain of cases
which goes from heat-waves, to epidemics, to earthquakes, 
to large-scale food crises. As in any chain, it is of
particular interest to consider more closely its two extremities.
Heat-waves in developed countries result in excess mortality%
\qfoot{In the sense of: [(observed deaths)-(deaths in normal
years)]/(deaths in normal years).}
of the order of 24\% (Rey et al. 2007, p. 536, Table 1) 
whereas in the food crises
that we are going to study mortality rates were increased by up to
140\%.  The successful identification of birth rate troughs
in the wake of heat-waves required a skillful and pioneering analysis
of {\it daily} birth data by Arnaud R\'egnier-Loilier (2010 a,b).
The reason why heat waves display only 
a faint Bertillon effect is not only due to low excess mortality
but also to it being concentrated in elderly people 
which makes it irrelevant for birth rates%
\qfoot{Even for persons in age of having children
the birth rate trough is only marginally due to
those who die. Most of the effect comes from the much larger
number of persons who are affected but do not die.}%
.
\qpar

Why did we say that heat-wave cases constitute the beginning
of the chain? One could of course consider diseases (e.g. the Lyme
disease) which have even lower mortality. However, as
the birth rate troughs of heat-waves
are already at the limit of what can be observed, the
Bertillon effect for Lyme disease would be inobservable.
It would be like a pulsar which is known to exist
but is too far away to be seen with an optical telescope.
\qpar

In the cases considered in the present study
the Bertillon effect is so massive that it
can be identified (and studied even at regional level) with {\it annual}
vital statistics. 

\qA{From well documented cases to uncertain situations}

Why did we choose to focus on the three cases selected? 
The main reason is that these cases are massive and
statistically fairly well documented.
\qpar
In contrast, there are many cases
of mass mortality for which there is considerable uncertainty.
Thanks to the death-birth relationship (to be stated below)
one is in good position
to throw new light on such episodes. How?
\qpar

Most countries, even those which do not have a reliable statistical
registration infrastructure, conduct censuses. Because this
is done periodically (e.g.
once in a decade) it does not require a permanent
organization.
As will be shown below, the distribution of the population
by age measured in a census gives a fairly
accurate picture of birth rate changes in the years
preceding the census. For instance, the census
of 1982 in China gives a good picture of the birth rate squeeze
that occurred in several provinces in 1961. 
Then, through the relationship
between birth troughs and death spikes one can get
an estimate of the mortality. Even though not perfect,
such estimates give at least a rough picture
of what happened.

\qA{Outline}
Our investigation will proceed through the following steps.
\qee{1} First, we present the three famine cases which will
be studied. Apart from giving some 
social and historical background information,
we will discuss the origin, reliability and accuracy of the 
birth and death data. 
\qee{2} Then, in each case we describe the
Bertillon death-birth connection.
\qee{3} In order to get a comparative view we bring together
the three Bertillon relationships and we compare them
with the results already obtained in Richmond et al. (2018b).
\qee{4} We show how to retrieve past annual birth rates
from a population pyramid.
\qee{5} Finally, in our conclusion, we develop two examples
in which mortality rates are derived from census data.

\vskip 8mm
\centerline{\color{magenta} \bf \Large PART 1: QUANTIFICATION 
OF THE BERTILLON EFFECT}
\vskip 2mm

\qI{Background information for the crises}

\qA{Causes of death}

First of all, we should explain why we prefer to
use the expression ``food crises'' rather than ``famines''.
The word ``famine'' elicits images of people starving to death.
Although, this may of course exist as documented by
impressive Internet pictures of children almost reduced to
their bones%
\qfoot{For instance the persons shown on the cover
of Davis (2000). Needless to say, the inclusion of pictures
which have such an emotional content also reveal something
about the agenda of the author.}%
, 
death by starvation represents only
a small fraction of the global death toll of food crises.
This can be illustrated by data from Finish and Indian sources.
\qbu The data given in Finland 1 (p. XXXIV) and Finland 2 (p. 421-422)
tell us that during the crisis of 1868 death by starvation
represented only 1.71\% of the deaths. The main cause of death
was typhus (43\%), followed by tuberculosis (5.84\%),
dysentery (5.70\%) and smallpox (3.02\%).
\qbu Data for the food crisis of 1942--1944 in India are given
in Maharatna (1992, p. 331). Death by starvation represented
2.1\% of the deaths. The main cause of death was fever due
to diseases (mainly malaria) which accounted for 34.8\% 
followed by scabies (18.4\%) and dysentery (10.9\%).
\qpar

In short, at symptoms level, food scarcity crises
have a close resemblance with epidemics.

\qA{Crisis of 1868 in Finland}

Immediate causes such as a rainy and cold
summers in 1866 and 1867 can be mentioned
but in order to get a real understanding the
crisis of 1868 should not be seen as an isolated event.
In fact, there had been similar crises in 1833 and 1856.
albeit of smaller magnitude (Flora et al. 1987, p, 24,51).
Whereas in the 1840s the average annual death rate was around
25\promille \ it reached 46\promille \ in 1833, 34\promille \ 
in 1856 and 78\promille \ in 1868.
\qpar
The main factor in this succession of crises 
was certainly the rapid growth of the population.
Between 1811 and 1865 it increased by 66\% which is
twice the 30\% increase in France in the same time interval
(Flora et al. 1987, p. 55-56). This 66\% increase
represented an average annual increase of 1.20\%
(against 0.55\% in France).
\qpar
Certainly
agricultural production did not increase at the same rate
which means that any bad harvest due to adverse weather
conditions would result in malnutrition or a more serious crisis.
The annual death rate of 78\promille \ reached in 1868 in Finland
was one of the highest ever observed anywhere.
\qpar

The severity of a food scarcity is best judged by the death
rate peak for this is a fairly intrinsic metric.
In contrast excess-deaths are very dependent
on the level of the baseline death rate chosen to
define normal conditions.\qL
It is true that at the level of Indian provinces
the death rate reached similar values: for
instance in 1900 it reached 88 in the ``Central Provinces''.
However, if nationwide
data were available they would certainly show
lower peak values for in such a large land
as India the crises were not completely synchronous.

\qA{Food crises in India}

Between 1860 and 1910 there were recurrent food crises
in India but most of them were limited to some parts
of the country. A brief description can be found in 
Roehner (1995, p.5-6).
\qpar

Can one explain the famines in India in the same way
as the famines in Finland that is to say by a rapid
increase of the population. The answer is no.
Assuming that the population estimate given
for 1820, namely 209 millions is indeed reliable,
one finds for the period 1820--1865
an average annual increase rate of
only 0.27\%. 
\qpar
However, if the population stagnated the amount
of food available in India may have decreased. 
One can mention three  reasons for a shrinking
food supply (Davis 2000, p. 59-66).
\qbu Between 1875 and 1900 Indian annual grain exports to
Great Britain increased 
from 3 to 10 million tons, equivalent to the annual nutrition 
of 25 million people. Davis does not give trade data for the
period before 1875.
However, it is known that
as a result of the repeal of the corn tariff around 1850
Britain's dependence on imported grain increased from 2\%  in the
the 1830s, to 24\% in the 1860s, and 65\% (for wheat) in the 1880s.
Thus, it is quite possible that the drain on Indian grain
started before 1875.
\qbu In Berar (central India) 
the acreage of cotton doubled between 1875 and 1900.
\qbu Although a colony, with regard to its budget India
was treated like an independent country. This meant
reimbursing the stockholders of the ``East India Company'',
paying the
costs of the 1857 revolt, supporting an army of about 100,000
that was employed not only in India but also
in foreign war theaters (e.g. Afghanistan, Tibet, Egypt, Ethiopia,
Sudan). Military expenditures alone represented about one third
of the budget. As a result, little funds were available
for agricultural improvement.
\qpar

Having said that, in India just as in Finland and in China,
the immediate causes of the crises were unfavorable weather
conditions, particularly drought which in Madras and Bombay provinces
resulted from a reduced monsoon rain season.

\qA{Food crisis in China, 1960--1961}

The explanation given for Finland also applies 
to China and in fact to even greater degree.
From 1949 to 1959 the Chinese population increased from 541
to 672 millions, an increase of 24.2\% which represented
an average annual increase of 2.20\%. 
This is 4 times faster than
the French rate of 0.55\% that we took as a yardstick 
in discussing the case of Finland. 
It is true that between 1953 and 1957 population and
foodgrain production increased at the same rate of 2.0\%
annually: the population from 588 to 646 millions
and the foodgrain production from 168 to 185 million tons.
However, after a bumper crop of 212 million tons in 1958
it declined due to bad weather conditions to the
point of being reduced in 1960 to its level
of 1957 at which time there had been about 50 million
fewer Chinese to feed (SNIE 1961, p.3). 
\qpar

At the same time the death rate fell from
20\promille \ to about 10\promille. It is the fall in the
death rate which brought about the rapid population increase
for the birth rate remained stable at around 40\promille.
Fig. 1 shows that even at its peak value in 1960 the death
rate reaches only 25\promille \ (Ren kou 1988, p. 268)
which is lower than the 
death rate in Bengal under normal conditions and only slightly
higher than the death rate in 1949. In short, instead of
people starving to death, one should rather think of
the situation as similar to the one of 1949%
\qfoot{According to the SNIE (1961. p.1) report discussed
in Appendix A: ``Widespread famine does not appear to be at hand but
in some provinces people are now on a bare subsistence diet.''}%
.
It is true that
the situation may have been more tragic in 
some specific provinces as for instance Anhui or Sichuan.
However, this was not really new for, as we will see below,
the best predictor for regional death rates
in 1960 was the situation one decade earlier.

\qA{Comparative graph}

Fig. 1 shows the changes in death rates in India
and China.
For the sake of clarity we did not represent the curve for
Finland on the same graph; however, it can be 
noted that 
the baseline death rate before the Great Famine of 1868
was around 28\promille \ and that the death rate peak reached
78\promille. 
In a general way, for a given
peak value, the lower the
baseline rate, the higher the excess mortality with respect
to this baseline. The fact that for China the baseline level
was 2 or 3 times lower than in India is the main reason
for a substantially higher excess death number.

%
\begin{figure}[htb]
\centerline{\psfig{width=14cm,figure=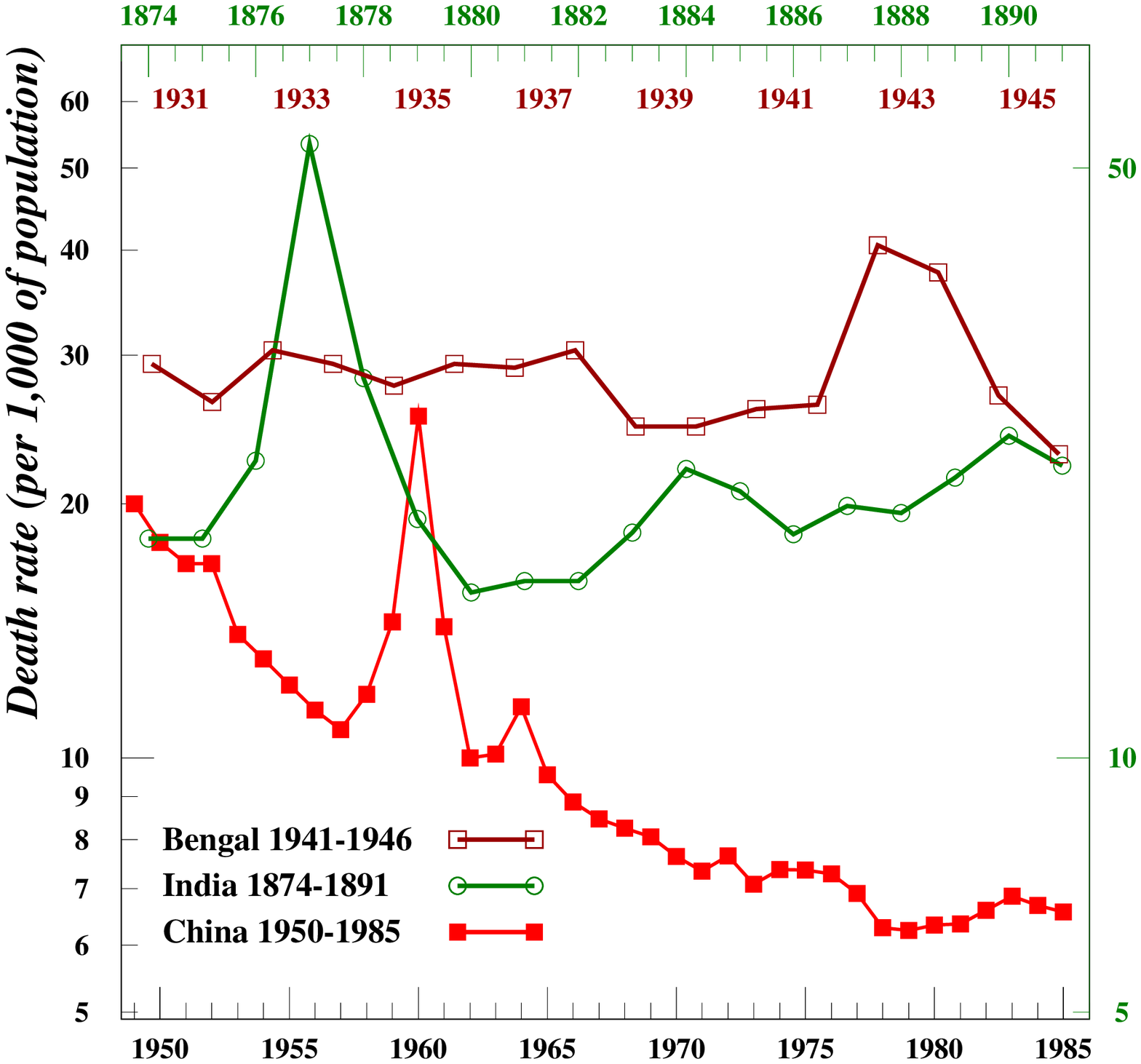}}
\qleg{Fig.\qhu 1\qhv Comparison of famines in India and
China.}
{The vertical scale is the same for the 3 curves. 
As there are no nationwide data for India the
curve corresponds to data for the province of
Madras in the south east of India.
}
{Sources: Maharatna (1992, p.55), Ren kou (1988, p.278).}
\end{figure}

In Fig. 1 it may appear surprising that in 1931--1940 the death rate
in Bengal
was about 1.5 times higher than the rate for 1874--875 in Madras.
The most likely  explanation is under-reporting.
The registration of deaths started
around 1870 and, as is often observed, the scope and completeness
of the
registration increased progressively in the course of following
decades. 
In the 1940s, the corrective factors used by various authors
still ranged from 1.32 to 1.70 (Maharatma p. 228). This suggests
that in the 1870s under-reporting may have been
more serious, may be by a factor 2.

\qI{The Bertillon death-birth relationship}

In each of the previous cases death and birth data are available
not only at national level but also at provincial level.
This will allow us to carry out a regression analysis in the
same spirit as in Richmond et al. (2018b). 
Here, however we will have
to work with yearly (not monthly) data. 
It is to maintain accuracy that
we limited ourselves to massive events.  
As in many
countries only annual data are available,
it is important to see whether or not in such cases
the Bertillon effect
can be analyzed in a meaningful way.

\qA{Bertillon effect in Finland: food crisis of 1868}

Fig. 2 shows the Bertillon effect for the 8 provinces
which composed Finland in the 1860s.
%
\begin{figure}[htb]
\centerline{\psfig{width=10cm,figure=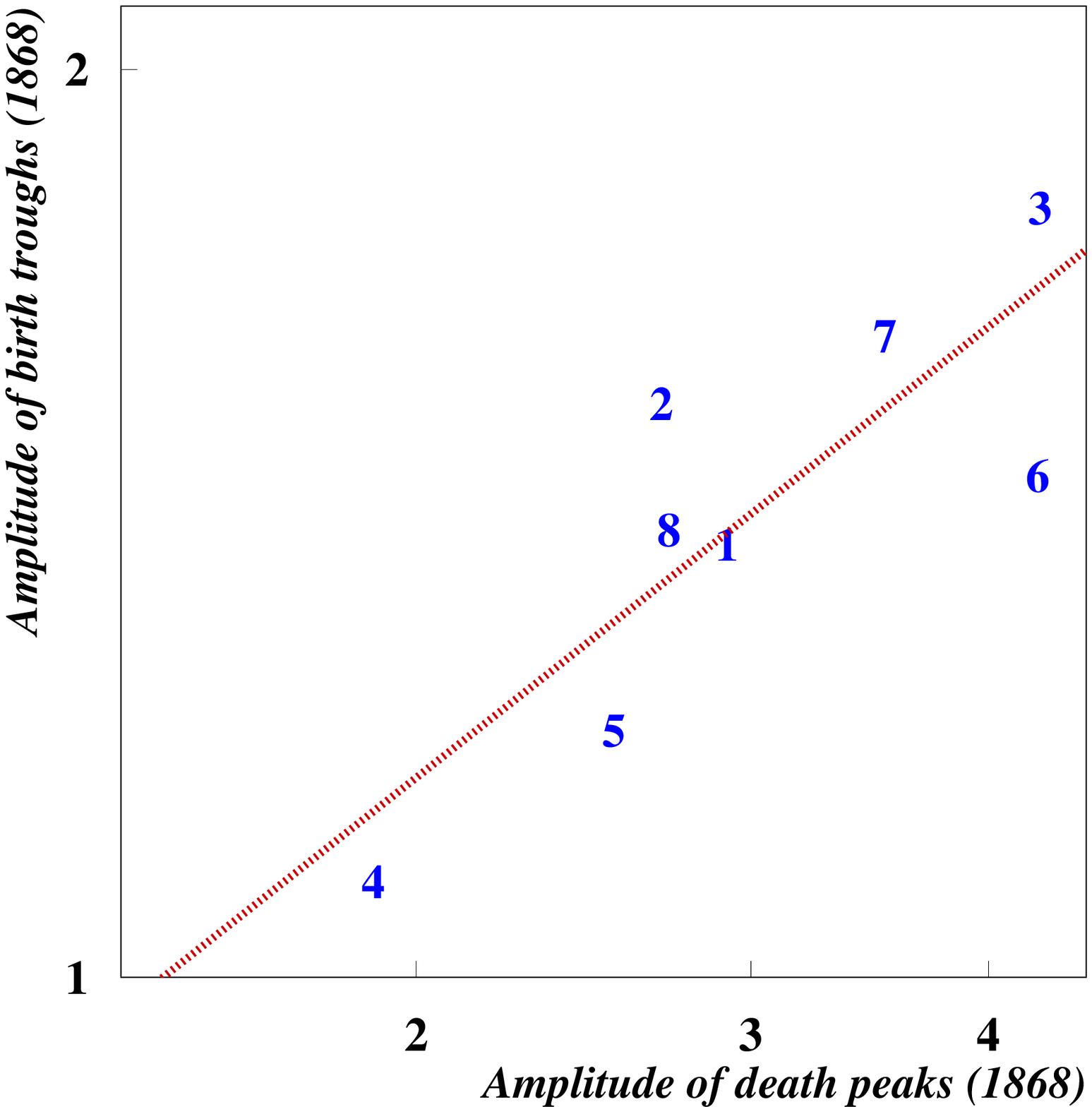}}
\qleg{Fig.\qhu 2\qhv Finland, 1868: relationship between the 
amplitudes of death spikes and birth troughs.}
{As the graph is
a log-log plot it means that: $ A_b=CA_d^{\alpha} $.
where $ \alpha=0.50\pm 0.26 $ and $ C=0.89\pm0.06 $  
(the error bars are for a confidence level of 0.95);
the coefficient of linear correlation is $ 0.83 $.
Each number corresponds to one of the 8 governments
(i.e. provinces) which composed Finland. Their names are
as follows: 1=Uudenmaan (S), 2=Turun ja Porin (SW),
3=H\"ameenlinnan (S),4=Wiipurin (SE), 5=Mikkelin (SE),
6=Kuopion (E), 7=Waasan (SW), 8=Oulun (N).}
{Sources: Finland (1871, p.34,XII), Finland (1902, p.274,397)}
\end{figure}

As monthly data are
available  for the whole country, we know that the 
death rate peak occurred in May 1868; therefore,
as expected, the center of the birth trough will be
in December 1868--January 1869. In other words the birth trough
will be split in two parts, one part in late 1868 and the
other in early 1869. However, because of the rebound effect%
\qfoot{The rebound effect occurs in the months immediately
 following the trough; it is
a compensating rise of the birth rate above normal
baseline level (see Maharatna 1992, p. 380 and Richmond et al. 2018a).
In other words
it is a return to equilibrium marked by an ``overshooting''
episode.} 
the annual birth number for 1869 is higher than the one
for 1868. That is why in Fig. 2 the vertical scale
displays births of 1868 and not 1869.

\qA{Birth troughs seen as a sensitive detector of population suffering}

At first sight it might seem that the absence of a death spike, i.e. 
$ A_d=1 $ means that the situation is normal and
should therefore be associated with
$ A_b=1 $. However, the fact that there are no excess-deaths
does not mean that the population does not suffer. A simple
illustration is a non fatal disease which nevertheless makes
people ill. The 2003 SARS epidemic in Hong Kong
discussed in Richmond et al. (2018a) was a situation
of this kind. Although not strictly equal to zero, the death
toll was only 40 per million. Nevertheless,  
the threat and disorganization due to the epidemic produced
an excess birth trough of 6\% ($ A_b=1.06 $). \qL
In the same line of thought it will be seen
in subsequent cases
that usually the birth rate starts to fall in an early stage
at a moment when no increase of the death rate can be detected.
In other words, birth
troughs are a detector of population suffering that is
more sensitive than $ A_d $.

\qA{Bertillon effect in India: time series of food crisis}

Here ``India'' refers to the British colony
before its division into Bangladesh, India and 
Pakistan. 
\qpar
As already observed there are no nation-wide data
for India in spite of the fact that some food crises
extended to several parts of the country. 
Instead one has data separately for several
provinces: Bombay, Central Provinces, Madras, Punjab,
United Provinces. Fig.3a,b displays the basic 
mechanism of food crises: as the death toll increases
the birth rate decreases. One may wonder why here, 
in contrast to Finland,
the birth trough occurs in the year following the 
death spike. Monthly data that are available in
some cases (Dyson 1991a,b, 1993 and Maharatna 1992, p. 235)
show that usually the maximum of the deaths occurred in
August at the beginning of the wet season. The reason for that
comes from the fact that the main cause of mortality is malaria
whose spread is favored by humidity.

%
\begin{figure}[htb]
\centerline{\psfig{width=17cm,figure=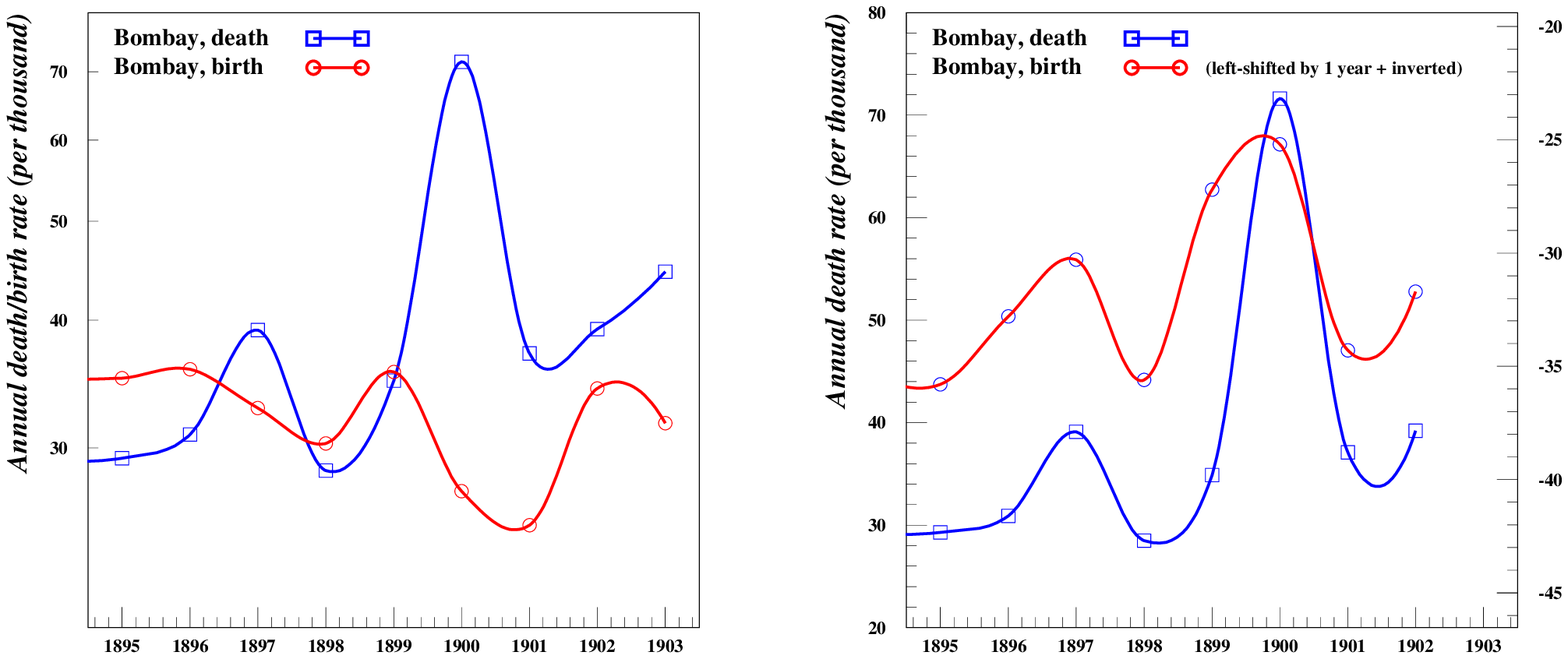}}
\qleg{Fig.\qhu 3a,b\qhv Bombay (1895-1903).}
{There are two food crises in the time interval 1895-1903:
first one of small magnitude in 1896-1897 and then the 
more serious crisis of 1899-1900. The right-hand graph
(where the birth curve was shifted and inverted)
makes manifest that both death spikes
gave rise to a birth trough in the following year.}
{Source: Maharatna (1992, p.55)}
\end{figure}
%
With deaths spiking in August instead of May as was the
case in Finland the effects of reduced conception will
be mostly visible in the following year.
\qpar

Fig. 3a and 3b show that the analysis performed
in Richmond et al. (2018a) on monthly data can
be repeated in a similar  way with annual data.

\qA{Bertillon effect in India: global death-birth relationship}

Next, we wish to discover the relationship 
existing between
the amplitudes of death spikes and birth troughs
of different crises. This is summarized in Fig. 4.
There is again a power-law relationship.

%
\begin{figure}[htb]
\centerline{\psfig{width=10cm,figure=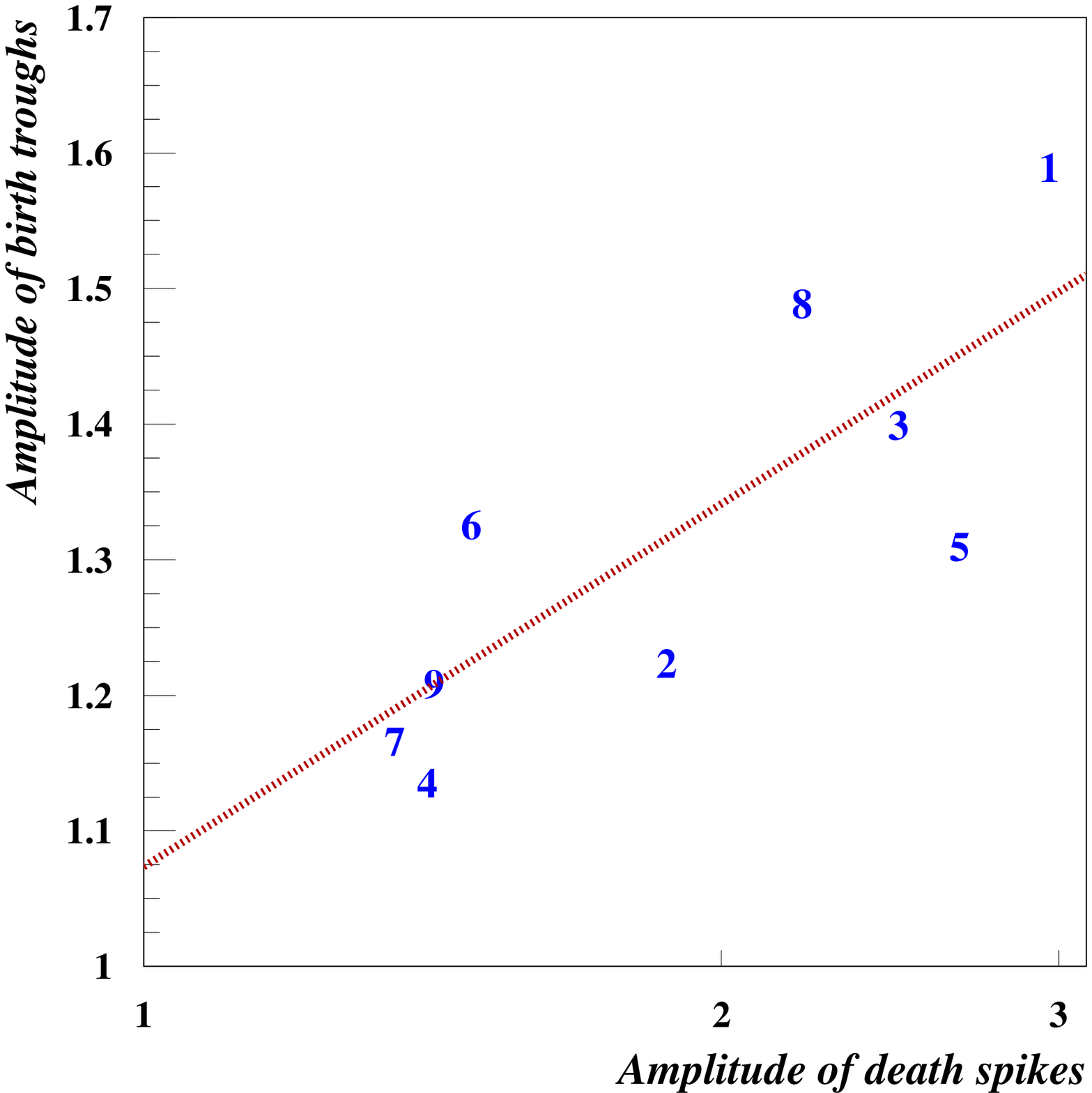}}
\qleg{Fig.\qhu 4\qhv India (1877-1908). Relationship between the 
amplitudes of death spikes and birth troughs.}
{The relationship is as follows:
$ A_b=CA_d^{\alpha} $.
where $ \alpha=0.30\pm 0.16 $ and $ C=1.01\pm0.06 $.
The coefficient of linear correlation is: $ 0.81 $. 
The correspondence between the numbers and the provinces is
as follows: 
1=Madras (1877), 2=Bombay (1877), 3=Bombay (1900),
4=Punjab (1900), 5=Central Provinces (1897),
6=United Provinces (1908),
7=Bombay (1897), 8=Central Provinces (1897), 9=Berar (1897).
}
{Source: Maharatna (1992, p.55).}
\end{figure}

\qA{Bertillon effect in China: time series of the 
food crisis of 1960-1961}

Of the three countries considered in this paper
it is for the case of China that we have the most
complete data set. Annual data are given for over
20 provinces from 1955 to 1985. In what follows we will
focus on a 10-year interval centered on the crisis
of 1960-1961.
\qpar

Fig. 4 shows typical graphs for 4 different provinces .
In a general way, the crisis was more serious in
South China than in North China and in the south the severity
increased from east to west.
Apart from Beijing, the three other selected provinces are 
all located at the same latitude (about 500 km south
of Shanghai): Fujian is on the Pacific coast, Hunan in central
China and Sichuan about 700km to the west of Hunan. 
%
\begin{figure}[htb]
\centerline{\psfig{width=14cm,figure=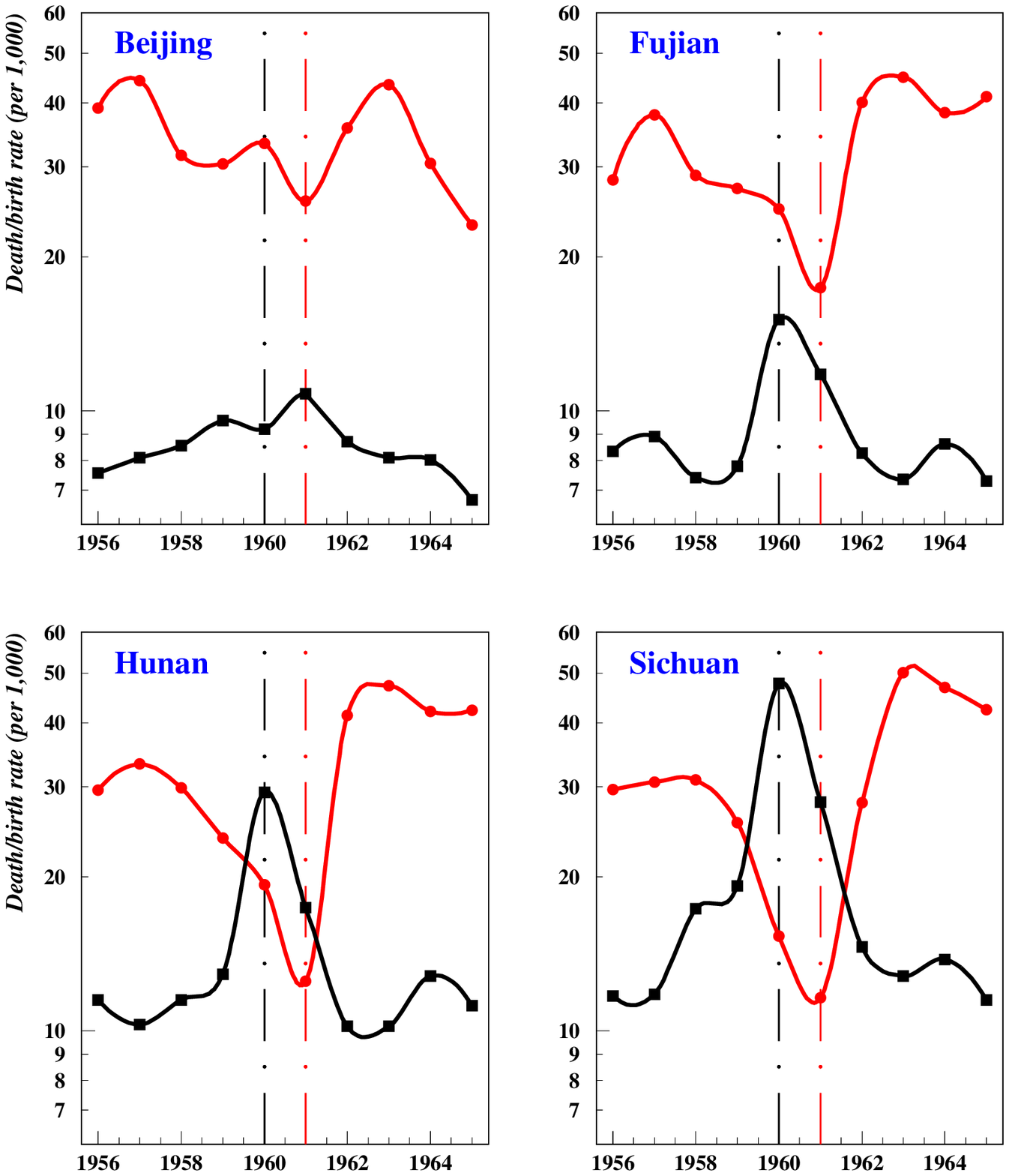}}
\qleg{Fig.\qhu 5\qhv Birth and death rates in 4 Chinese
provinces.}
{The curve of deaths is in black with filled squares;
the curve of birth is in red with filled circles.
The graphs are ranked by order of increasing severity.
It can be seen that the fall of the birth rates in 1957, 1958
provide
so to say early warnings of an impeding crisis.
The fact that the minima of the birth troughs occur in 1861
shows that the mortality rates certainly peaked in the second
half of 1960. }
{Source: Ren kou (1988)}
\end{figure}
%
The graphs show very clearly
two characteristics of the Bertillon effect.
\qbu The one year time lag between death peaks and birth troughs
\qbu The birth rate rebounds in the two or three years
following the troughs.
\qpar

In the next subsection,
in order to find out the relationship
between the amplitudes of death peaks and birth troughs,
we extend this analysis to 22 provinces

\qA{Bertillon effect in China: global death-birth relationship}

Fig. 6 shows the relationship between death peaks and
birth troughs. As in Finland and India it can be described
by a power law. The exponent is about the same as in Finland.

%
\begin{figure}[htb]
\centerline{\psfig{width=12cm,figure=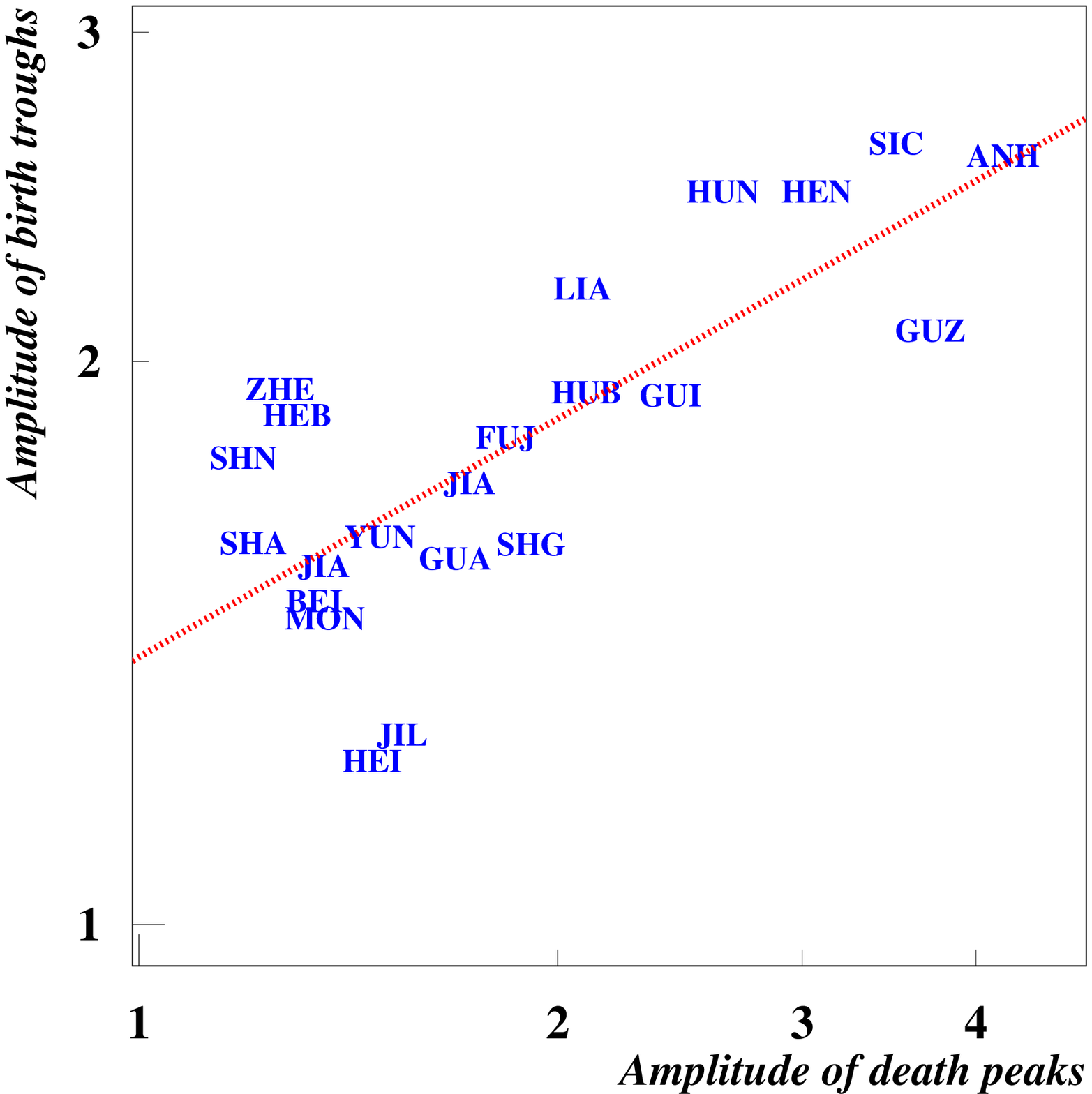}}
\qleg{Fig.\qhu 6\qhv China 1959-1961. Relationship between the 
amplitudes of death spikes and birth troughs.}
{The relationship between the amplitudes of the death and
birth rates is as follows:
$ A_b=CA_d^{\alpha} $.
where $ \alpha=0.42\pm 0.16 $ and $ C=1.38\pm 0.06 $.
The coefficient of linear correlation is: $ 0.76 $. 
The labels are the first three letters of the names of the
provinces except for the following whose first three letters
would have been identical: 
SHA=Shanxi, SHN=Shanghai, SHG=Shandong, 
GUA=Guangdong, GUI=Guanxi, GUZ=Guizhou.}
{Source: Ren kou (1988)}
\end{figure}

\qA{Structural fragility}

One may wonder what made the crisis more serious 
in some provinces than in others. We have already mentioned
that globally the north was less affected than the south and
the east less than the west. However, besides these fairly loose
observations there is another which is both simpler and more
revealing. It consists in the fact that the provinces
which have the highest death rate peaks in 1960 are those which
already had the highest death rate in the normal years before
the crisis. This is illustrated graphically in Fig. 7.

%
\begin{figure}[htb]
\centerline{\psfig{width=12cm,figure=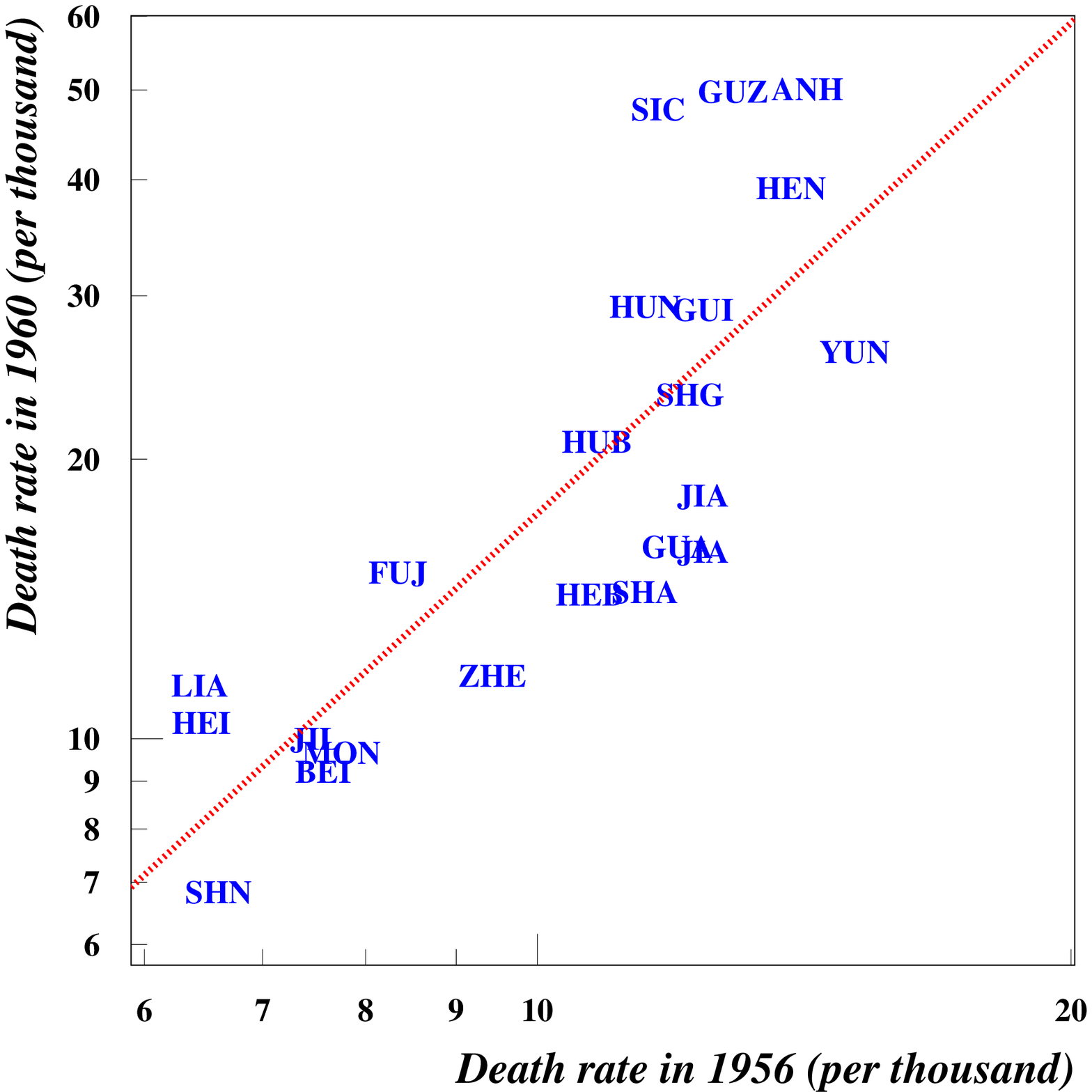}}
\qleg{Fig.\qhu 7\qhv Relationship between the 
death rates in 1956 and in 1960 (food crisis).}
{The coefficient of linear correlation is: $ 0.81 $. 
The labels are the same as in Fig.5.}
{Source: Ren kou (1988)}
\end{figure}

Is there a similar connection between normal and peak death
rates in Finland and in India?
It is impossible to say because in those cases
the number of separate provinces for which data are available
is too small.
\qpar

What interpretation can one give of the effect shown in Fig. 6?
The higher death rates in normal times suggest structural
fragility factors. To get better insight 
it is useful to compare extreme cases.
On one end we consider the northern provinces of
Heilongjiang and Jilin where the crisis was subdued while on the
other end we consider Anhui and Sichuan where it was
very severe.
\qpar

In Table 1 we computed the population growth between 1954 and
1959. A much faster growth in Anhui-Sichuan would suggest
that, as was indeed the case for the whole country, the
increase of food supply could not match the population
growth. However, the data in Table 1 do not
show a faster population increase in Anhui-Sichuan.
As a matter of fact, had it existed such a faster growth
would not explain that the death rate in Anhui-Sichuan
was already higher in 1956. This fact rather suggests that
there was a structural fragility that was already present
in 1956 but had an aggravated effect in 1960.

%
\begin{table}[htb]

\small
\centerline{\bf Table 1: Regional fragility}

\vskip 5mm
\hrule
\vskip 0.7mm
\hrule
\vskip 0.5mm
$$ \matrix{
&  \hbox{Population} & \hbox{Population} & \hbox{Percent} \hfill &
  \hbox{\color{red} Population}\hfill&
\hbox{Death} \hfill & \hbox{Death} \hfill\cr
&  & & \hbox{increase} \hfill & \hbox{\color{red} density}\hfill&
\hbox{rate} \hfill & \hbox{rate} \hfill\cr
\qtb
&  1954 & 1959 &  \hfill & \color{red} 1954\hfill&
1954 \hfill & 1960 \hfill\cr
\qtb
&  (10^6) & (10^6) &  \hfill & \hbox{(per sq-km)}\hfill&
\hbox{(per 1,000)} \hfill & \hbox{(per 1,000)} \hfill\cr
\noalign{\hrule}
\qth
\hbox{Jilin} \hfill & 11.7  & 13.2  & 13\% \hfill & \color{red}60\hfill&
10.3 \hfill & 10.1 \hfill\cr
\hbox{Heilongjiang} \hfill & 12.7  & 16.9 &33\%  \hfill &\color{red} 37\hfill&
10.5 \hfill &  10.5\hfill\cr
\hbox{} \hfill &   &  &  \hfill & \hfill&
 \hfill &  \hfill\cr
\hbox{Anhui} \hfill & 31.7  &34.4  & 8.5\% \hfill & \color{red}224\hfill&
16.4 \hfill &  50.2\hfill\cr
\qtb
\hbox{Sichuan} \hfill &  64.4 & 73.7 &  14\% \hfill & \color{red}134\hfill&
 15.4\hfill &  47.8 \hfill\cr
\noalign{\hrule}
} $$
\vskip 1.5mm
Notes: Jilin and Heilongjiang were little affected by the crisis
whereas Anhui and Sichuan were the two provinces where the
crisis was the most severe. It seems that population density
is the main factor which can explain this difference; in
Anhui-Sichuan it was on average 3.7 times larger than in
Jilin-Heilongjiang. This created a permanent fragility and
sensitivity to weather conditions which is revealed by the
high death rates even before the crisis.
\qL
Source: Ren kou (1988)
\vskip 2mm
\hrule
\vskip 0.7mm
\hrule
\end{table}
%

The data in Table 1 show that the population density was
far higher in Anhui-Sichuan than in Jilin-Heilongjiang.
The difference is all the more striking because
in contrast to Jilin-Heilongjiang
Sichuan had at that time almost no industry.  
\qpar

Another contributing factor was the regional
prevalence differential of infectious diseases.
It is well known that malnutrition reduces the ability
of the organism to fight infection.
For tuberculosis in 1990 the prevalence was 147 per
100,000 in the East, 198 in the Center and 216 in the
west (China Tuberculosis Control Collaboration 2004, p.421)
\qpar

We will not develop this analysis further for it would lead
us too far away from the main purpose
of our paper which to focus on general rules.
In the next subsection we compare the regression lines in
Finland, India and China.

\qA{Comparison of the regression parameters}

When the three regression lines are drawn on the same graph
one can compare their slopes (given by the exponent $ \alpha $)
and their levels (given by the coefficient $ C $). 
The level is given by the birth trough amplitude for a 
given death peak amplitude. One should realize that
this depend very much upon how the birth deficit is
divided between the two years. Thus, for a death peak in
May (as in Finland) the birth deficit will be divided
almost equally between 1868 and 1869. On the contrary,
if deaths had peaked in November 1868 then 
the whole birth deficit would occur in 1869.
In other words, to make this comparison significant
one would need monthly data. 
\qpar

With respect to the slopes we observe that they are
fairly similar for Finland and China but 50\% smaller
for India. Regarding the value of $ \alpha $ in India one can observe
that in this country
the data cover several food crises in successive years whereas
in Finland and China they cover a single crisis.
However, it is not clear why this should lead to a
lower $ \alpha $.
\qpar

For the influenza epidemic of October 1918 in the United States
the slope was (Richmond et al. 2018b): 
$ \alpha=0.19\pm 0.1 $ and $ C=1.28\pm 0.22 $.
Here, a lower value of $ \alpha $ makes sense.
To make the argument simpler, let us consider as an approximation
that $ C=1 $; this means that in all cases the regression line
starts from the point $ (1,1) $.
Then the question becomes: ``For a
given death peak how many people will suffer to the point
of reducing conceptions?'' It makes sense to observe that 
during a food crisis more people do suffer, and suffer more
severely, than during an influenza epidemic. An obvious reason
is that the influenza epidemic was much shorter than 
any of the food-crises that we considered.
It lasted only about one month, from 15 October to 15 November 1918
whereas the food crises lasted more than one year.
\qpar

In the last section before the conclusion
we will show how the
death-birth relationship can be used as an exploration 
tool for historical mass mortality events.

\vskip 8mm
\centerline{\color{magenta} \Large \bf
PART 2: MASS MORTALITY EPISODES EXPLORED}
\vskip 2mm
\centerline{\color{magenta} \Large \bf THROUGH THE BERTILLON EFFECT}
\vskip 4mm

\qA{Why are mortality statistics often uncertain?}

The production of vital 
statistics comprises two fairly different tasks.
\qbu In order to register individual births and deaths
one needs a network extending to hospitals and doctors of
the whole country. This is a challenging
task as can be seen from the fact that in a vast country like
the United States it took several decades to extend the
registration
network to all states. The task was completed only around 1930.
Even once established, registration networks
may be overwhelmed during mass mortality episodes.
\qbu The second task is the organization of
censuses. Although by no means an easy task, the organization of
censuses does not require a permanent network
extending to the whole country. In a time span 
of 3 or 4 months the same team of census officers can
move from region to region until the whole country has
been covered. \qL
The fact that the organization of a census is a much less
demanding task than daily registration is demonstrated
by the observation that the first US census took place about
one century before the national registration network was completed. 

\qA{How can one derive annual birth numbers from census records?}

In principle, a census does not give any information about
births or deaths but it gives the age of all people and
from these data one can derive approximate birth data.
This is illustrated in Fig. 8a,b.

%
\begin{figure}[htb]
\centerline{\psfig{width=17cm,figure=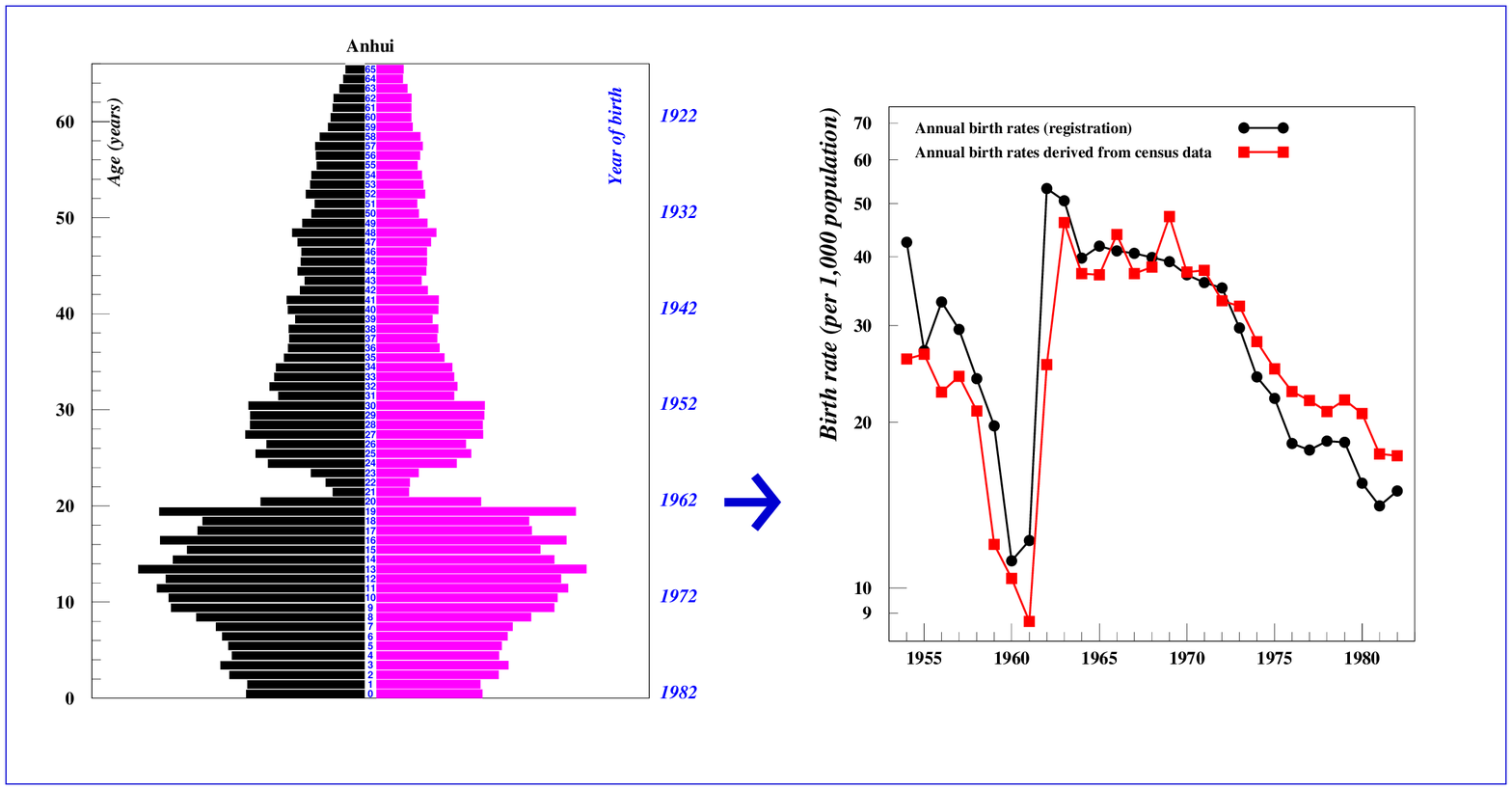}}
\qleg{Fig.\qhu 8a,b\qhv Derivation of birth data from
a population pyramid.}
{{\bf (a)} The huge indentation is due to the birth
deficit in 1960-1961.
The population pyramid of Anhui province
was built with the data of the census of 1982.
Each step represents an age group born in a single year.
Males are on the left and females on the right.
The birth rate trough corresponds to people who
were $ 1982-1961=21 $ year old at the moment of the census.
People older than 21 were born before 1961 and people
younger than 21 were born after 1961. 
In this way one can reconstitute approximate birth data
for the whole period from 1955 to 1982.
{\bf (b)} Graph 7b is for males; its horizontal axis shows
the years of birth as derived from the ages recorded in 
the census of 1982.
It shows a fairly good
agreement between the real birth data and the approximate
birth data derived from the population pyramid.}
{Sources: Population pyramid: IPUMS. Birth data: Ren kou (1988).}
\end{figure}

In what sense are the data derived from the 
population pyramid approximate birth numbers? In principle
the size $ b'(1982) $ of the age group $ 0-1 $
of the population  pyramid should
be equal to the number $ b(1982) $ of births in 1982. The
fact that in graph 7b the two points are not identical
is because we have been using a 1\% sample of
the census of 1982. This
gives an idea of the statistical fluctuations. More generally,
the individuals aged $ x $ in 1982 were born in $ y_b=1982-x $.
\qpar
For instance,
the individuals born in $ y_b=1970 $ will be 12 year old in 1982.
Naturally, during these 12 years some of these children 
died or moved from Anhui to another place
(whether in China or abroad); conversely
some children may have moved into Anhui. If these  
population movements are important the size of the 12-year
age group will have little to do with the 
number of births in 1970. On the contrary, if there were
few population movements a comparison of the numbers $ b'(x) $
and $ b(x) $ will tell us how well the registration
system was working. Generally speaking, the registration 
of births is more reliable than the registration of
deaths for at least two reasons.
\qbu When funeral services are overwhelmed by the number
of deceased people it may happen that they are buried by
relatives or neighbors without being registered, especially
in times of epidemics.
On the contrary, as there are no birth rate spikes the
birth registration services will not get overwhelmed.
In addition
the registration of newborns does not need to be done immediately
after birth, it can be done at any time.
\qbu Politically, dead people are usually a more sensitive
matter than newborns. The authorities may wish to
publish under-estimated death numbers%
\qfoot{For instance, according to
 ``R\'esum\'e r\'etrospectif (1907, p. 368), the death rate in
Ireland in the period 1864--1870 was 16.6\promille \ which was
the lowest rate among all European countries. 
It was lower than in
England and Wales (22.5), Sweden (20.2), Denmark (19.9) or
Prussia (27.0). Most likely such a small rate resulted
from under-reporting.}%
.

\qA{Preliminary test of the methodology on influenza deaths in
Pennsylvania}

First of all, before using it as an exploration tool,
we wish to test the accuracy of the methodology on a case
in which the death toll is known. Why did we select 
for our test the
influenza epidemic in Pennsylvania?
The impact of the disease was particularly severe in
Pennsylvania. The amplitude of the death peak of 1918
(with respect to the average of the 1917 and 1919 numbers) was
1.56 in Pennsylvania but only 1.35 for the 
whole country (or more precisely for all death registration
states). 
\qpar
The challenge now is to see if we can derive the death peak
amplitude solely from the age-group data given by the census of
1930. For that purpose we need to go through the following steps.
\qee{1} First, as was already done in the previous subsection,
we derive from the age-groups given by the census of 1930
proxy birth numbers. To distinguish these proxy birth numbers
from the real birth numbers (given by the registration network)
they will be called {\it census birth numbers}. This
step is done in Fig. 9a,b. In contrast with Fig. 8a,b
here there is no huge trough. Before doing the test we could
not know whether or not the small trough of 1918 would be
covered by the background noise. Fortunately, it turns out
that it can be identified. Its amplitude is 1.09.

%
\begin{figure}[htb]
\centerline{\psfig{width=17cm,figure=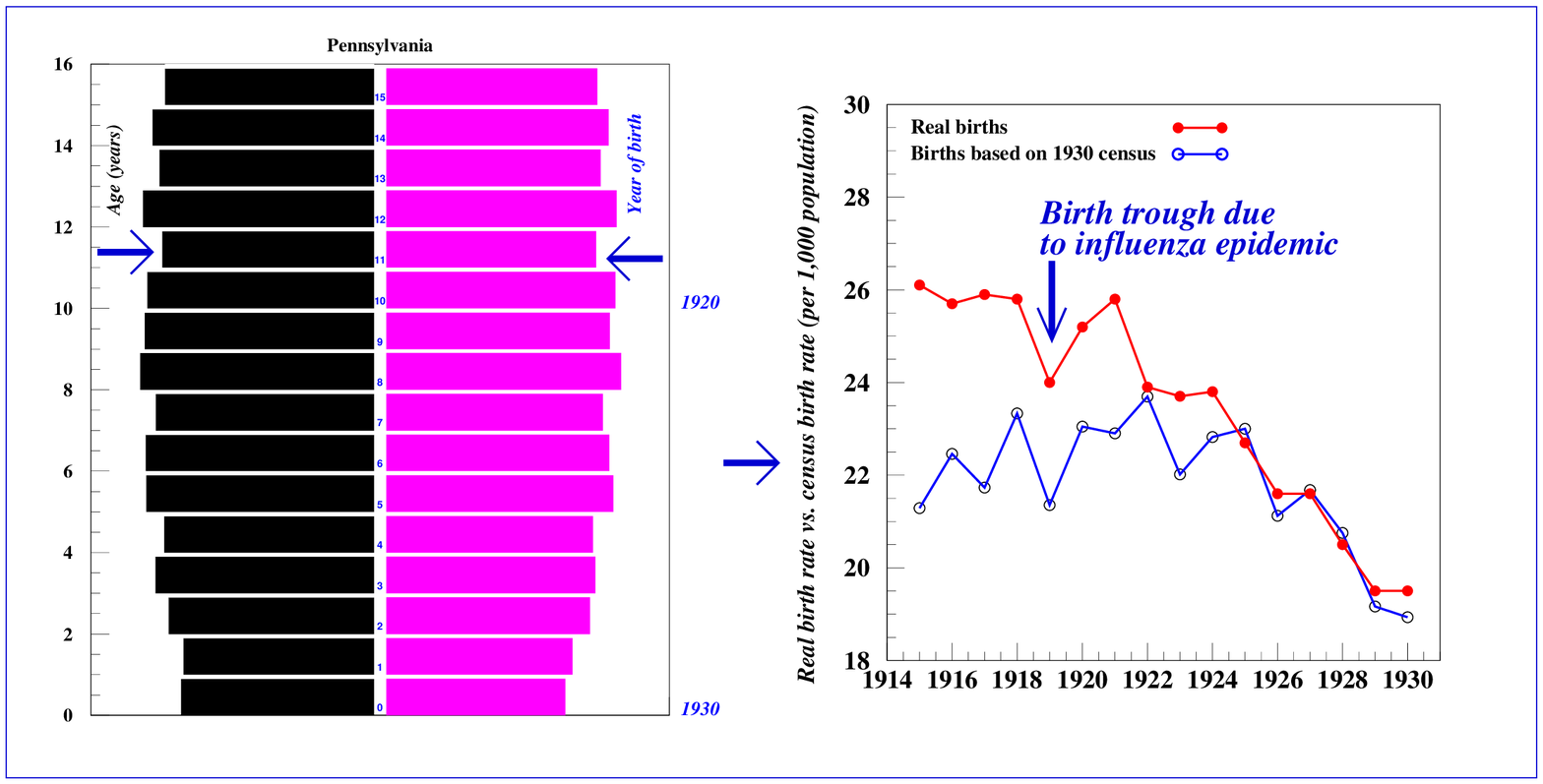}}
\qleg{Fig.\qhu 9a,b\qhv Derivation of birth data from
the population pyramid of Pennsylvania.}
{{\bf (a)} The population pyramid was truncated to the
ages that are required for our exploration of the
birth trough of 1919 (indicated by  arrows).
{\bf (b)}  In 1930 the two data points should be identical.
The small difference seen in the graph is due to the
statistical fluctuations resulting from the fact that
the population pyramid was derived from a 5\% sample of the census
of 1930. As one moves back in time it is of course natural
that the distance between the two curves increases because
the people born in these years appear in 1930 as
older age-groups.}
{Sources: Population pyramid: IPUMS, Birth data: Linder et al. 1947, 
p.666-667.}
\end{figure}

\qee{2} The second step is to use the death-birth relationship
to derive the amplitude of the death peak.
As the level of the regression line is not well defined
(as was explained eralier it depends upon the month of the
death spike) we take again $ C=1 $ for the multiplicative
constant. Thus, using the value of $ \alpha $ given previously,
one gets:
$$ A_b = A_d^{0.19} \quad \Rightarrow
A_d=A_b^{1/0.19}=1.09^{5.26}=1.57 $$

\qee{3} Now, in order to derive the death rate in 1918
from the amplitude of the death peak, we need an 
estimate for the death rate in normal times. 
Although in some cases this death rate may be known,
most often it is not. In the two cases considered below the
normal death rates are not known for the simple reason
that there was no registration network. In such a cas
one takes the death rate in a region that is similar in
terms of socio-economic conditions. For
instance one may take the
death rate in 1915 in Sweden which is 14.7\promille \
(Flora et al. 1987, p.73).
 Here, as the death rate in Pennsylvania is in fact well
known (for in 1915 it was already a death registration state)
we can check whether the Swedish proxy is reasonable.
It is indeed for in Pennsylvania in 1915 the death rate
was 13.2\promille. \qL
Thus, the predicted death rate
in 1918 becomes: $ \mu=1.57\times 14.7=23.1 $.\qL
Compared with the real value of 18.1\promille\ there is
a difference of 28\%. This could seem high,
but one should remember that in such matters 
the estimates by various authors often differ by 100\% or
more. In the first case considered below, no estimate
whatsoever was available and in the second case
the official estimate may have been highly exagerated.

\qA{Georgia during the American Civil War}

As an illustration of the kind of situation for which
the birth-death method may be useful
consider the state of Georgia in 1864, the year of the Civil
War during which General Sherman led Union troops through
Georgia on what is called the ``March to the Sea''(Nov.--Dec 1864)
which resulted in great damages.

%
\begin{figure}[htb]
\centerline{\psfig{width=15cm,figure=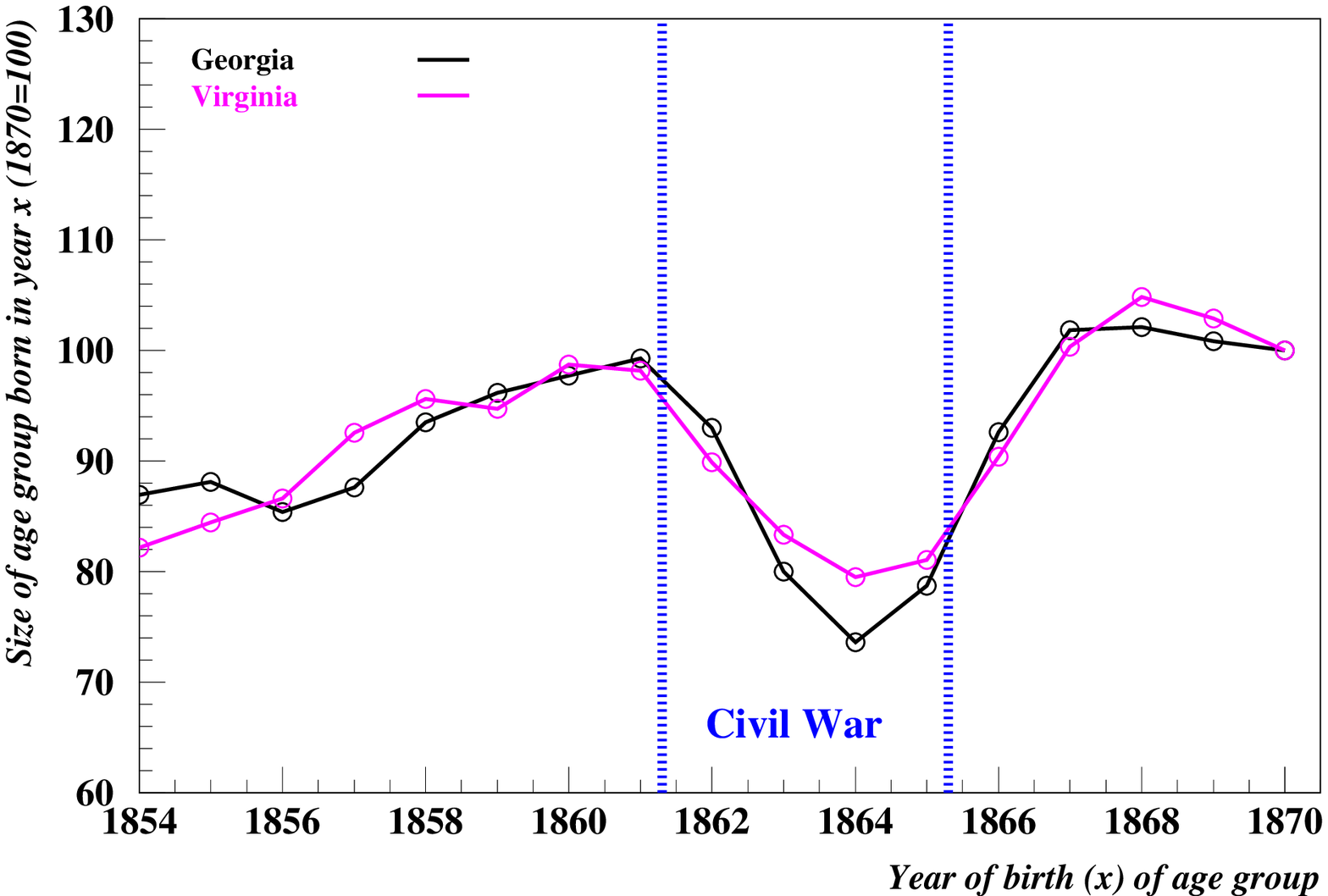}}
\qleg{Fig.\qhu 10\qhv : Reduction in birth rates during the
Civil War in Georgia and Virginia}
{The horizontal axis shows the year of birth as derived
from the age at the census of 1870.
The fact that in 1861 and 1870 the levels are nearly
the same shows that there was no permanent emigration
or immigration between the year of birth and the census year.}
{Source: 1\% sample of the census of 1870 from the database
of IPUMS, USA.}
\end{figure}

At that time Georgia was not a registration state which
means that no death data were recorded. As a result, one
has no idea of the civilian death toll of the ``March to  
the Sea''. However, there was a census in 1870
through which one can measure the birth numbers in earlier years.
The population pyramid shows that there was a birth trough of
amplitude 1.33 and of width 3 years (Fig. 10).
Applying the relationship
between birth troughs and death peaks leads to a death
peak amplitude $ A_d=(1/C)A_b^{1/\alpha} $. For the sake of
simplicity we take $ C=1,\ \alpha=0.5 $ which will give a fairly
conservative death estimate: $ A_d=1.33^2=1.77 $. 
For the average death rate in 1861--1870 time we take the rate of 
Sweden namely 20\promille \ (R\'esum\'e r\'etrospectif 1907, p. 368).
Thus the peak rate in 1864 was: $ 1.77\times 20=35 $\promille . 
In 1860 the population of Georgia was about 1 million; this
leads to an excess-death toll of $ 15\times 1000=15,000 $.
There may be a additional excess-deaths in 1861--1863 but
as we do not know the width of the death peak (usually it is
more narrow than the birth trough) we will not try to
estimate them.
\qpar

The figure of 15,000 excess-deaths should be seen as a tentative
estimate because it is not obvious that the death-birth relationship
derived from food crises can also be used for a war time situation.
During wars, the fact that husbands and wives are separated 
may inflate the birth troughs to an extent which may critically
depend upon how frequently servicemen can return home.
If the birth troughs of Fig.8 are inflated through the separation
effect, the figure of 15,000 will be an over-estimate.

\qA{Measles epidemic in Fiji Islands in 1875}

It is claimed that the measles epidemic of 1875 in Fiji
claimed one third of the population which was assumed
to have been 150,000 before the epidemic (McArthur 1967, p.8).
Note that both the population and the death toll are
merely estimates made by western visitors%
\qfoot{For instance, the population estimates
published by different visitors range from 100,000 to 300,000.}%
.
%
\begin{figure}[htb]
\centerline{\psfig{width=8cm,figure=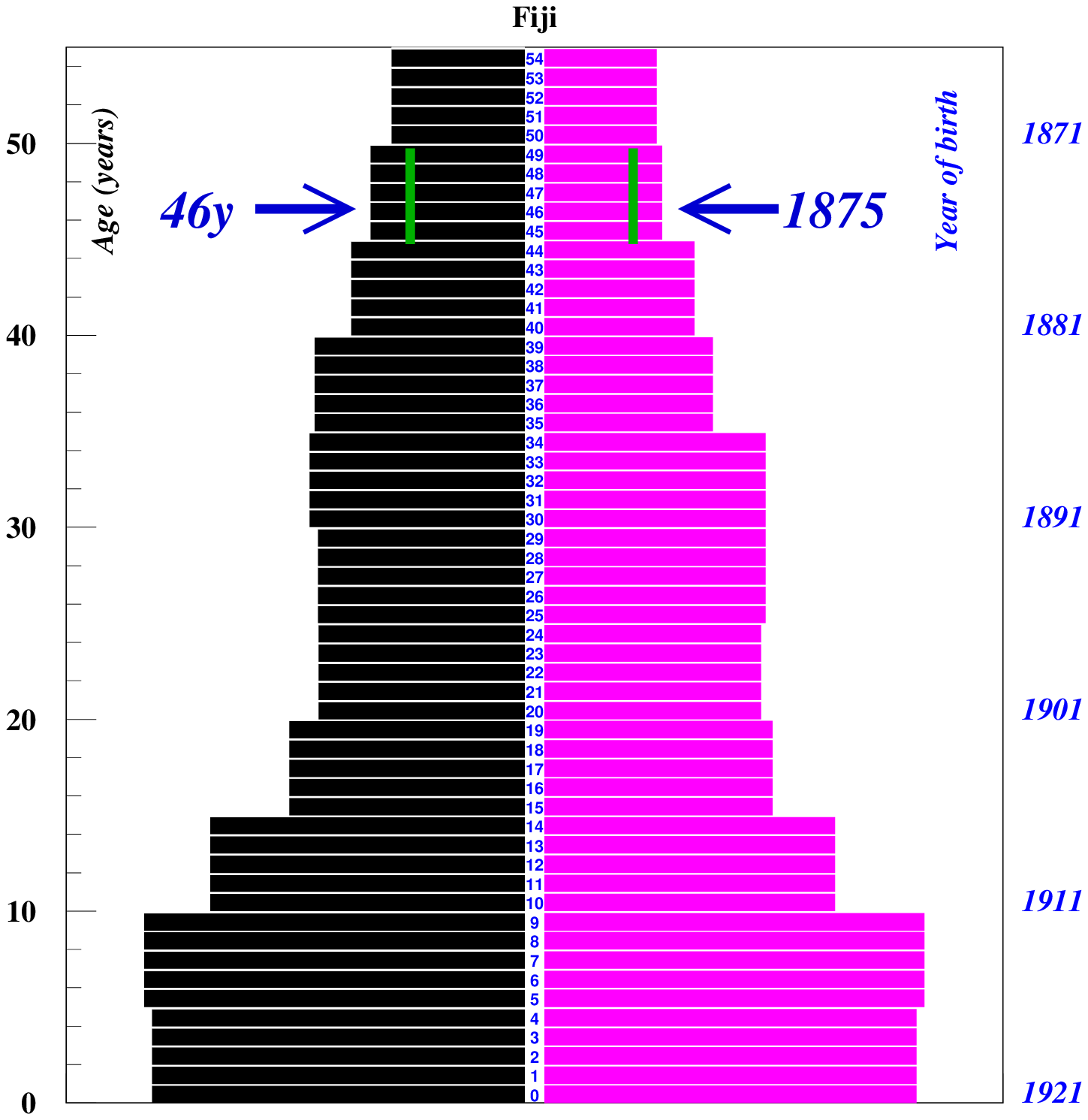}}
\qleg{Fig.\qhu 11\qhv Was there a devastating measles
epidemic in Fiji in 1875?}
{The ages given on the left-hand side correspond to
the age-groups in 1921. One expects a birth trough in 1875.
In the census of 1921 this trough will be seen as a 
reduced size of the age-group that is 49 year old
resulting in a reduced 49 year old age-group. 
Based on a death-birth relationship with an exponent $ \alpha=0.5 $,
one would expect a reduction of the corresponding 
5-year age group as indicated in green.}
{Source: Fiji census of 1921, the first
in which ages were recorded, cited in McArthur (1967, p.38).}
\end{figure}

This sounds
somewhat surprising when compared to the death toll of
the influenza epidemic of 1918 which was only 5\%. 
This last figure is probably more reliable than the
one for 1875 because in the meanwhile the procedure
for births and deaths registration was improved by the British
administration. Fig. 11 is based on the census of 1921.
This was not the first one but in the censuses of 1879,1881,
1891, 1901 and 1911 only 4 age groups were distinguished,
namely: children, youths, adults and aged. \qL
Even without
doing any further calculation a comparison with Anhui
(where the death toll was 5\%) shows that
the 33\% claim seems dubious. It is true that instead 
of annual data here we have only 5-year averages but
in Anhui the contraction would still be clearly visible
on 5-year averages. More specifically,
a calculation involves the following steps.
(i) Computation of the amplitude of the death spike based on
the 33\% death toll.
(ii) Computation of the amplitude of the birth trough
based on the death-birth relationship; for 
$ \alpha $  we selected $ \alpha=0.5 $ but taking another
value (e.g. $ \alpha=0.3 $) would not make a great difference.
(iii) Finally, from the amplitude of the birth trough one derives
the expected contraction of the 5-year age group (shown in green in
Fig. 11).
\qpar

One may wonder why British authorities were favorable to the
thesis of a big drop in the Fijian population%
\qfoot{In British accounts the gravity of this epidemic is
heavily emphasized (see McArthur 1967, p.8). The following
excerpt is from a report by a commission in 1893.
In 1875 ``the people were estimated 
to number about 150,000, and it is recorded, 
probably with fair exactitude,
that 40,000 died from measles during the epidemic which overran
the whole archipelago in the space of 4 months.
Whether the Fijians who survived have had their
stamina permanently lowered can only be a matter for conjecture.''}%
. 
One reason which comes to mind is that in 1879 started
a massive transfer of Indian indentured workers to Fiji
which had the purpose of providing
the manpower needed on sugar cane plantations. By 1946 the
Indian community had developed to the point of being
larger than the native Fijian population.

\qI{Conclusion}

\qA{Outline of the exploration of food crises}

In his seminal paper Jacques Bertillon used weekly and monthly
birth and death data (Bertillon 1892). We wished to see to
what extent his analysis can also be performed using annual
data for in many countries only annual vital statistics are 
publicly available.
It was shown that the 
death-birth effect can be analyzed in a significant way
provided one considers
events marked by large-scale mortality. The annual death
birth data were found to be compatible with the 9-month
time lag revealed by weekly and monthly data
(see Fig. 3.5). 
\qpar

By observing regional
death peaks and birth troughs we found a power law relationship
between their respective amplitudes in confirmation of
a similar connection already observed in Richmond et al. (2018b).

\qA{Agenda for future explorations of mass mortality episodes}

The spirit of a death registration network is to
go bottom up. Starting from block level one should move up
to county, state and nation level. At each step the
death numbers should be aggregrated until one gets to 
a total number for the whole  country.
Any death estimate made at the macro-level without
being supported by appropriate data at micro-level
should be considered as suspicious. 
Unfortunately in many cases of interest the data 
that would permit a bottom up procedure are 
simply not available. 
\qpar

The methodology developed in the previous section
allows us to use censuses made decades after the
mass mortality. For instance, in the case of the Fiji
Islands we have been using a census made 4 decades after
the epidemic of 1875. We believe that this procedure
can bring new light in many dubious estimates.

\appendix

\qI{Appendix A: Specific features of the crisis in
  China}

It seems that in the course of the past two decades 
the factual description of the food crisis of 
1960-1961 has progressively
been replaced by accounts based on ill-founded
stories.  The Wikipedia articles entitled:
``Great China famine'' and ``History of agriculture in 
the People's Republic of China'' illustrate this tendency.
The second one contents claims not based on any reference
or outright mistakes%
\qfoot{Such as the claim that 
China did not import grains before 1962
which, as will be seen below, is not correct.}%
.
Science is not only about building nice
models, first and foremost it means starting from the
right facts.
In order to come back to a more
scientific description, it is important to collect
as many hard data as possible. By hard data we mean
data that can be checked in some way. 
This is for instance the case of grain trade
data because such figures are published by the two trade partners,
first as exports, then as imports. It is the purpose
of the present appendix to present a number of useful data.

\qA{Parallel with the food crisis and meteorological situation of
  1878}

An article entitled ``La famine en Chine'' [Famine in China]
was published in 1880 in a French scientific journal (Margoll\'e
1880) which describes a meteorological situation very 
similar to the one in 1959--1960, namely a severe drought in north
China over 3 years together with floodings in south China.
In this article, the two phenomena
are explained by the existence of a strong high altitude
air circulation from west to east which prevented
the steady humid flow originating in the south from
reaching the north. Instead the accumulated humidity would
eventually be released in central and south China and
provoke the floodings. 
\qpar

The severity of the famine
was described as follows.
``Between 1876 and 1878, a lethal drought-famine struck the five
northern provinces of Shanxi, Henan, Shandong, Hebei, and
Shaanxi. By the time the rains returned late in 1878,
an estimated 9 to 13 million of the affected area's total
population of about 108 million people had perished
(Legge 1878, Edgerton-Tarpley 2008). The publication of 1878 by the
``Committee of the China Famine Relief Fund'' shows that 
this famine triggered
a mobilization in western countries. 
Relief operations opened a road to the action of western missionaries.
The famine took place 14
years after western powers helped the imperial government to defeat the
Taiping rebellion.

\qA{Excess deaths}

When the death rate of 1958 is taken as the baseline
the peak of 1958--1961 gives an excess death toll
of 14 millions, an impressive figure%
\qfoot{If one takes as baseline the level of 1950 the
excess death toll is about 7 millions which is of the
order of magnitude of the recurrent famines that
occurred in the 1920s and 1930s.}%
.
As in Finland the crisis
was a two step process: the fast population
growth made the country vulnerable to adverse weather conditions;
then in to successive years 1959 and 1960
the country was hit by severe droughts. We will also try to
assess the role of the ``Great Leap'' policy. 

\qA{Wheat imports}

The situation was aggravated by the difficulties and delays
in importing  grains. 
The USSR 
was unable to export grains to China%
\qfoot{The following articles published 
in the ``New York Times''  provide some useful glimpses
into the situation.
\qbu 10 May 1959: Drought perils wheat crop
in North China. 
\qbu 19 August 1959: China is believed to 
be coping successfully with one of 
the worst droughts in mainland China in several decades.
\qbu 27 August 1959: China announced that its 1958 production figures
issued earlier
this year, had been overstated. It also reduced output goals set for
1959.
\qbu 12 February 1960: Taiwan buys for \$5.2 million (about 0.1
million tons) of US wheat.
\qbu 5 July 1960: The Chinese government spurs efforts to rush
irrigation equipment into drought-struck north east provinces
to save wheat crop.
\qbu 7 August 1960: The Chinese government is resorting to
extraordinary measures to produce food. Millions of people
are shifted
from nonagricultural jobs to fight shortage.
\qbu 16 October 1960: Several million tons of wild plants
have been collected in north China against losses
caused by crop failure, the Peiping radio has reported.
\qbu 23 January 1961: In 1960 China and the
Soviet Union had an ``unsatisfactory agricultural output''.
\qbu 4 May 1961: An agreement was signed
between China and Canada for the sale of 6 million tons
of grains (at a cost of \$362 millions)
over a period of 2.5 years. 
\qbu 13 September 1961: Canada is also
exporting grains to the Soviet Union and Poland.
\qbu 10 March 1977: After three years of reducing its 
wheat imports, 
China is once again diping deeply into its foreign currency 
reserves to feed its people. Total wheat imports reached 
\$550 million. [At a price of \$55/ton this represents
10 million tons, i.e. some 6.5\% of Chinese consumption.
Without such massive imports the situation might have
become quite as dramatic as in 1960.]
\qbu 20 March 1977: China's economic performance last year was the
worst since the Cultural Revolution in the late 1960's, with the
growth rates of industrial and agricultural production well below
those of 1975. [According to government data, the real GDP
growth rate was 8.7\% in 1975 and -1.6\% in 1976.]
}%
. 
A table of the SNIE (1961) reports that there
were no Soviet grain exports to China.
Soviet-China relations 
already started to sour in 1960. In July-August 1960
the USSR abruptly withdrew nearly all of the
2,500 Soviet industrial technicians present in China;
moreover, a disruption in oil delivery  created a shortage
of petroleum products in late 1960
(SNIE 1961, p. 4).
\qpar

After 1950
the United States had put in place a drastic trade embargo
which, through the CHICOM committee, involved also US allies.
In 1959 the US had a large surplus of wheat.
An article of 8 November 1959 in the New York Times is
entitled ``Wheat surplus is a big headache''.
It says that at the start of the crop year the surplus
stood at 1.27 billion bushels (on the basis of one bushel
of wheat  weighing 30kg this quantity represented 38 million
tons) enough to meet US domestic
needs for two years. Nevertheless,
US exports to China were practically nil between 1950 and 1970.
\qpar

Led by its fairly independent Prime Minister John
Diefenbaker, Canada was willing to brave the embargo. It
sold a small amount of wheat to China 
in 1958 and opened discussions with the 
US to get permission to sell more. 
Indeed, in 1961 Canada sold about 1 million tons
of wheat and barley to China. This represented one third
of total Chinese imports of foodgrains in 1961 (Lu 1997, p.22)
but less than 1\% of 
the total production of around 170 million tons according
to SNIE (1961, p.3). It was both too little and too late%
\qfoot{A discrepancy can be noted between
Lu (1997, p.22) and SNIE (1961, p.6) regarding the total
imports of grain in 1960: the first source gives 5.81 million
tons whereas the second gives 2.81 million tons.
The reason of this difference
is that the second number is based on contracts
which had been concluded by April 1961 when this report was
published. This means that there were additional contracts and 
deliveries between April and December 1961, probably from Burma
and Malaysia.}%
.
In summary, it can be said that
through its own embargo and its influence on
other western grain exporting countries, the United States
bears at least partial
responsibility in the crisis of 1960.

\qA{Conflicting accounts}

Most present-day accounts blame
the ``Great Leap Forward'' as the main cause of
the food crisis. However, a US intelligence
report (SNIE 1961) which was published in April 1961
gives another perspective. It
is of interest for several reasons.
\qee{1} Classified as ``Secret'', this joint report
of the intelligence organizations of the Army, Navy, Air Force,
State Department 
and of the Central Intelligence Agency, was not 
 destined to be published. Therefore, one
cannot claim that it was part of a public relations operation
and does not reflect the opinions of its authors.
As a matter of fact, it was declassified only in
December 1996.
\qee{2} As it was published in April 1961 it is possible
to check whether its predictions were confirmed by
subsequent events. This test supports indeed
the perspective which is presented.
\qee{3} The SNIE report contrasts with present-day 
mainstream accounts in several important ways.
For instance, it describes in
detail the measures already taken in 1960; they show that,
contrary to the claim made in the Wikipedia articles
mentioned above, 
the leadership was well informed about the 
reality of the situation.

\qA{Was the ``Great Leap Forward'' an economic failure?}

Nowadays it is a standard and almost self-evident belief
that  the ``Great Leap
Forward'' (1957--1960) was a technical and economic failure%
\qfoot{Though also held in China, this view is more
based on ideological reasons developed during the 
Deng period than on hard facts.}%
.
It is indeed quite likely that it put 
additional stress on country side
people, but was it an  economic 
failure? Although only of marginal
importance for the purpose of the
present paper,
from a scientific perspective the question certainly deserves
to be raised.
\qpar

The following facts can be mentioned.
\qbu A short Internet search shows that at least 5 large
dams were built in those years. Here is the list.\qL
(i) 1957--1960, Sanmenxia Dam, Henan/Shanxi, 106m, 16.2 cubic-km \qL
(ii) 1958--1962, Xinfengjiang Dam, Guangdong, 105 m, 13.9 cubic-km \qL
(iii) 1958--1962, Zhexi Dam, Hunan Province, 104 m, 3.65 cubic-km \qL
(iv) 1958--1981, Chengbihe Dam, Guangxi, 70 m, 1.12 cubic-km 
\qpar

In 1964, Ren\'e Dumont, a French expert in agricultural economics
wrote the following (Dumont 1964, p. 393, our translation): 
``Between 1955 and 1964
I observed the most extraordinary transformation of the agricultural
landscape. When one flies over China from Hanoi to
Beijing one sees that the regions to the south of the Yangtze
are now covered with canals, levees and dikes''. 
Naturally, the other side of the coin
was that for this kind of work men
were often employed far away from their villages which disrupted
family life and reduced conceptions.
This may have amplified
the Bertillon birth effect. It is a fact that birth reductions
were much more pronounced south of the Yangtze than in northern  
provinces.
\qbu The SNIE (1961) report gives growth estimates for
industrial production and GDP which are 
summarized in Table A1. The forecast for 1961 was made under the 
assumption that Soviet technicians would not come back to China;
otherwise it would have been higher.
%
\begin{table}[htb]

\small
\centerline{\bf Table A1: Estimates for industrial
production and GDP growth during the ``Great Leap Forward''}

\vskip 5mm
\hrule
\vskip 0.7mm
\hrule
\vskip 0.5mm
$$ \matrix{
&  1958 & 1959 & 1960 & 1961 & \hbox{\color{red} Average}\cr
\qtb
&   &  &  & \hbox{(forecast)}&\cr
\noalign{\hrule}
\qth
\hbox{Industrial production} \hfill &  & 33\% & 16\% & 12\% & \color{red}20\%\cr
\qtb
\hbox{GDP} \hfill & 18\% & 12\% & 8\% & & \color{red} 13\%\cr
\noalign{\hrule}
} $$
\vskip 1.5mm
Notes: It is often said that the ``Great Leap Forward'' was
a technical failure but the present
estimates made by the US intelligence community tell another
story. It is true that such a rapid growth
was not sustainable but the spirit of doing things 
notably faster than
elsewhere is still present in China nowadays. 
The construction
of the high speed rail track between Beijing and Shanghai 
was completed in 3-4 years, whereas in France the construction
of a similar line between Lille and Montpellier (almost
the same distance) took about 15 years.
\qL
Source: SNIE (1961, p.2,5) [Special National Intelligence
Estimate].
\vskip 2mm
\hrule
\vskip 0.7mm
\hrule
\end{table}
%

Quite understandably the report does not say how these
figures were computed but the fact that making such estimates was one
of the main duties of US intelligence agencies suggests
that they had means to do that reliably.
Here again, the fact that the report was not destined to be made
public is important because otherwise the publication
of estimates could be a way to influence the public
opinion. 
\qpar

Despite the slowdown of 1960-1961
the average rates remain impressive even for a fast-growth economy
like China. They justify the SNIE's observation that
``Peiping recognized that it could not continue the breakneck
industrialization tempo of 1958--1959''.
\qpar

According to government statistics released later on,
there was a decline in the GDP in 1962--1963. It is difficult 
to separate the after-effect of the food crisis from
the consequences of the departure of the Soviet technicians,
the shutdown (or reduction) of Soviet oil supply and other
adverse conditions connected with the end of the Soviet
economic cooperation. The end of Soviet assistance was particularly
critical in the face of the continued
western and Japanese trade embargo at least
until 1971.

\vskip 5mm

{\bf \large References}

\qparr
Bertillon (L.-A.) 1872: Article ``Mariage'' in the
Dictionnaire Encyclop\'edique des Sciences M\'edicales,
[Encyclopedic Dictionary of the Medical Sciences].
2nd series, Vol. 5, p.7-52.\qL
[In this study the author emphasizes the protective
role of marriage not only with respect to suicide
but also with respect to general mortality.]

\qparr
Bertillon (J.) 1892: La grippe \`a Paris et dans quelques
autres villes de France et de l'\'etranger en 1889--1890
[The influenza epidemic in Paris and in some other cities
in western Europe].
In : Annuaire statistique de la ville de Paris pour l'ann\'ee 1890, 
p. 101-132.
Imprimerie Municipale, Paris.\qL
[Available on Internet, for instance at the following address:\qL
{\small 
http://www.biusante.parisdescartes.fr/histoire/medica/resultats/index.php?do=livre\&cote=20955}]

\qparr
China Tuberculosis Control Collaboration 2004: The effect of
tuberculosis control in China. Lancet 364,417-422.

\qparr
Davis (M.) 2000: Late Victorian holocausts: El Ni\~no famines and the
making of the Third World. Verso, London.

\qparr
Dumont (R.) 1964: Les communes populaires rurales chinoises
[The people's communes in rural China].
Politique \'Etrang\`ere 29,4,380-397.\qL
[available online]

\qparr
Durkheim (E.) 1895: Les r\`egles de la m\'ethode sociologique.
F\'elix Alcan, Paris. Translated into English under the title:
``The rules of sociological method''.\qL
[The French and English versions are freely available on Internet.]

\qparr
Espinas (A.) 1878, 1935: Des soci\'et\'es animales [on animal
societies]. Thesis of the university of Paris. Republished
in 1935, F\'elix Alcan, Paris.\qL
[To measure the novelty of this study one should remember that
the author was in fact a sociologist.]

\qparr
Dyson (T.) 1991a: On the demography of South Asian famines. Part 1.
Population Studies 45,1,5-25.

\qparr
Dyson (T.) 1991b: On the demography of South Asian famines. Part 2.
Population Studies 45,2,279-297.

\qparr
Dyson (T.) 1993: Demographic responses in South Asia.
Institut of Development Studies 24,4,17-26.

\qparr
Edgerton-Tarpley (K.) 2008: Tears from iron: cultural responses to famine in
nineteenth-century China. The north China famine of 1876--1879.
University of California Press, Berkeley.

\qparr
Finland 1871. \qL
The Finnish title of this publication is:
Suomenmaan Virallinen Tilasto VI [Official Statistics of
Finland], Helsinki. \qL
[Available on line (April 2018): http://www.doria.fi/handle/10024/67301]

\qparr
Finland 1902. \qL
The French subtitle of this official
publication is: \'El\'ements d\'emographiques principaux de la
Finlande pour les ann\'ees 1750-1890, II: Mouvement de la population.
[Vital statistics of Finland for the years 1750--1890.]
The Finnish title is:  Suomen [Finland] V\"aest\"otilastosta
[demographical elements] vuosilta [years] 1750-1890, II:
V\"aest\"on [population] muutokset [changes]. Helsinki 1902. (526 p.)\qL
[Available on line (April 2018):
http://www.doria.fi/handle/10024/67344]

\qparr
Flora (P.), Kraus (F.), Pfenning (W.) 1987: State, economy, and
society in western Europe 1815-1975. A data handbook in two
volumes. Volume 2: The growth of industrial societies and
capitalist economies.
Macmillan Press, London.

\qparr
IPUMS: Integrated Public Use Microdata Series. 
Minesota Population Center,
University of Minnesota.\qL
[The IPUMS database consists 
of microdata samples from the censuses of the US
and a number of other countries. Most often the 
samples are 1\% or 5\% samples but for the US there are a few
100\% samples.] 

\qparr
Legge (J.) 1878: The famine in China. Illustrations by a native artist
with a translation of the Chinese text. Published by C. Kegan Paul
for the
``Committee of the China Famine Relief Fund'', London.

\qparr
Linder (F.E.), Grove (R.D.) 1947: Vital statistics rates in the
United States 1900--1940. United States Public Health Service,
Washington.

\qparr
Lu (F.) 1997: China's grain trade policy and its domestic grain
economy. Working paper No E1997002.
Peking University and Hong Kong university of Science and Technology.
[The paper contains a table which gives Chinese grain imports
and exports from 1953 to 1992. Grain imports were negligible until
1961 when they jumped to 5.81 million tons.] 

\qparr
Maharatna (A.) 1992: The demography of Indian famines: a
historical perspective. Thesis, 
London School of Economics and Political Science.
University of London.\qL
[In 1996 the thesis was published by Oxford University Press (Dehli)
under the title:
``The demography of famines: an Indian historical perspective.'']

\qparr
Margoll\'e (E.) 1880: La famine en Chine. 
La Nature, Revue des Sciences p. 314-315.

\qparr
R\'egnier-Loilier (A.) 2010a: \'Evolution de la saisonnalit\'e des
naissances en France de 1975 \`a nos jours [Changes in the seasonal
birth pattern in France from 1975 to 2006].
Population, 65,1,147-189.\ql
[This paper contains a section about the effect of heat waves.]

\qparr
R\'egnier-Loilier (A.) 2010b: \'Evolution de la r\'epartition
des naissances dans l'ann\'ee en France
[Changes in the seasonal birth pattern in France].
Actes du XVe colloque national de d\'emographie [Proceedings
of the 15th national conference on demography, 24-26 May 2010]
Published by the ``Conf\'erence Universitaire de d\'emographie
et d'\'etude des populations''.\qL
[Fig. 1 of this paper gives the pattern of monthly variations
of conceptions in France from the 17th to the 20th century;
it is based on Dup\^aquier (1976).]

\qparr
Ren kou 1988. \qL
The complete title has the following pinyin transcription:\qL
Ren kou tong ji zi liao hui bian.
Zhong hua ren min gong he guo. 
[Population statistics data compilation, 1949-1985. 
People's Republic of China.]. Published in 1988.\qL
[This official demographic report (1030 p.) gives annual birth and death
rates for the whole country and separately for each province.
It contains also population data by age and province
based on the three censuses of 1953, 1964 and 1982.
So far, we could not find an English translation but in 
fact the Chinese version can be used fairly easily;
one needs to know only a few key-words.]

\qparr
R\'esum\'e r\'etrospectif 1907: Statistique Internationale du
mouvement de la population d'apr\`es les registres d'\'etat civil.
depuis l'origine des statistiques de l'\'etat civil jusqu'en 1905.
[International vital statistics from the start of official
registration until 1905].
Imprimerie Nationale, Paris.

\qparr
Rey (G.), Fouillet (A.), Jougla (E.), H\'emon (D.) 2007:
Vagues de chaleur, fluctuations ordinaires des temp\'erature
et mortalit\'e en France depuis 1971.
Population 62,3,533-564.

\qparr
Richmond (P.), Roehner (B.M.) 2018a:
Coupling between death spikes and birth troughs.
Part 1: Evidence.
Physica A 506,97-111.

\qparr
Richmond (P.), Roehner (B.M.) 2018b:
Coupling between death spikes and birth troughs.
Part 2: Comparative analysis of salient features.
Physica A 506,88-96.

\qparr
Roehner (B.M.) 1995: Theory of markets. Trade and
space-time patterns of price fluctuations. Springer,
Heidelberg.

\qparr
Roehner (B.M.) 2010: How can population pyramids be used
to explore the past? Asia Pacific Center for Theoretical
Physics (APCTP) Bulletin, No 25-26, p. 13-25,
January--December 2010. 

\qparr
SNIE [Special National Intelligence Estimate number 13-61] 1961:
The economic situation in Communist China. 
Submitted by the Director of Central Intelligence on
4 April 1961. Declassified on 24 December 1996.\qL
[available on line]

\qparr
Viswanathan (G.M.), Luz (M. G. E. da), Raposo (E.P.), Stanley (H.E.)
2011: The Physics of foraging.
An introduction to random searches and biological encounters.
Cambridge University Press, Cambridge (UK).

\end{document}